\documentclass[useAMS,usenatbib]{mn2e}

\usepackage[active]{srcltx}
\usepackage{graphicx}
\usepackage{txfonts}
\usepackage{natbib}
\usepackage{multirow}
\usepackage{longtable}
\usepackage[applemac]{inputenc}
\usepackage{rotating}
\usepackage{tablefootnote}
\usepackage{url}
\usepackage[multiple]{footmisc}

\usepackage{color}
\usepackage{soul}
\definecolor{jaune}{rgb}{1.0, 1.0, 0.0}

\newcommand{\ra}{$R_{\rm A}$}
\newcommand{\rk}{$R_{\rm K}$}
\newcommand{\vinf}{$v_{\infty}$}
\newcommand{\teff}{$T_{\rm eff}$}
\newcommand{\mdot}{$\dot{M}$}

\newcommand{\bz}{\ensuremath{\langle B_z\rangle}}
\newcommand{\nz}{\ensuremath{\langle N_z\rangle}}

\newcommand{\kms}{km\,s$^{-1}\,$}
\newcommand{\vsini}{$v\sin i\,$}

\def\gtrsim{\mathrel{\hbox{\rlap{\hbox{\lower4pt\hbox{$\sim$}}}\hbox{$>$}}}}
\def\ltsim{\mathrel{\hbox{\rlap{\hbox{\lower4pt\hbox{$\sim$}}}\hbox{$<$}}}}

\title[HD\,164492C]{HD\,164492C: a rapidly-rotating, H$\alpha$-bright, magnetic early B star associated with a 12.5d spectroscopic binary\thanks{Based on observations obtained at the Canada-France-Hawaii Telescope (CFHT) which is operated by the National Research Council of Canada, the Institut National des Sciences de l'Univers (INSU) of the Centre National de la Recherche Scientifique of France, and the University of Hawaii. Based on observations made with ESO Telescopes at the La Silla Paranal Observatory under programme IDs 191.D-0255, 091.C- 0713 and 093.D-0267.}}

\author[G.A. Wade et al.]{G.A. Wade\thanks{E-mail: wade-g@rmc.ca}$^1$, M. Shultz$^{1,2}$, J. Sikora$^{1,2}$, M.-\'E. Bernier$^1$, Th. Rivinius$^3$, E. Alecian$^{4}$,
\newauthor{V. Petit$^{5}$, J.H. Grunhut$^6$ and the BinaMIcS collaboration}
\\
$^{1}$Department of Physics, Royal Military College of Canada, PO Box 17000 Station Forces, Kingston, ON, Canada K7K 0C6 \\
$^{2}$Department of Physics, Engineering Physics and Astronomy, Queen's University, 99 University Avenue, Kingston, ON K7L 3N6, Canada\\
$^{3}$European Organisation for Astronomical Research in the Southern Hemisphere, Casilla 19001, Santiago, Chile\\
$^{4}$Universit\'e Grenoble Alpes, CNRS, IPAG, F-38000 Grenoble, France\\
$^5$Department of Physics and Space Sciences, Florida Institute of Technology, Melbourne, FL, 32904, USA\\
$^6$Dunlap Institute for Astronomy \& Astrophysics, University of Toronto, 50 St. George Street, Toronto, ON, Canada M5S 3H4\\
}

\begin{document}

\date{Accepted . Received , in original form }

\pagerange{\pageref{firstpage}--\pageref{lastpage}} \pubyear{2002}

\maketitle

\label{firstpage}

\begin{abstract}
We employ high resolution spectroscopy and spectropolarimetry to derive the physical properties and magnetic characteristics of the multiple system HD\,164492C, located in the young open cluster M20. The spectrum reveals evidence of 3 components: a broad-lined early B star (HD\,164492C1), a narrow-lined early B star (HD\,164492C2), and a late B star (HD\,164492C3). Components C2 and C3 exhibit significant ($>100$~km/s) bulk radial velocity variations with a period of $12.5351(7)$~d that we attribute to eccentric binary motion around a common centre-of-mass. Component C1 exhibits no detectable radial velocity variations. Using constraints derived from modeling the orbit of the C2+C3 binary and from synthesis of the combined spectrum, we determine the approximate physical characteristics of the components. We conclude that a coherent evolutionary solution consistent with the published age of M20 implies a distance to M20 of $0.9\pm 0.2$~kpc, corresponding to the smallest published values. We confirm the detection of a strong magnetic field in the combined spectrum. The  field is clearly associated with the broad-lined C1 component of the system. Repeated measurement of the longitudinal magnetic field allows the derivation of the rotation period of the magnetic star, $P_{\rm rot}=1.36986(6)$~d. We derive the star's magnetic geometry, finding $i=63\pm 6\degr$, $\beta=33\pm 6\degr$ and a dipole polar strength $B_{\rm d}=7.9^{+1.2}_{-1.0}$~kG. Strong emission - varying according to the magnetic period - is detected in the H$\alpha$ profile. This is consistent with the presence of a centrifugal magnetosphere surrounding the rapidly rotating magnetic C1 component. 
 \end{abstract}

\begin{keywords}
Stars : rotation -- Stars: massive -- Instrumentation : spectropolarimetry -- Stars: magnetic fields
\end{keywords}

\section{Introduction}

The Trifid nebula (M20) is a very young and active site of star formation, with an estimated age of $(0.1-1)\times 10^6$~yr \citep{1998Sci...282..462C,2008hsf2.book..509R,2011ApJ...738...46T}. The distance is uncertain, with values from 1.7-2.84 kpc discussed in the recent literature \citep{1985ApJ...294..578L,2008hsf2.book..509R,2011A&A...527A.141C}, although even smaller distances (under 1 kpc) have been proposed \citep{1999A&AS..134..129K}. 

HD\,164492C is one of several bright, early-type stars composing the central asterism of M20\footnote{This composite image situates the HD\,164492 grouping within the larger nebular complex: {www.robgendlerastropics.com/M20-Subaru-HST.html}.}. Fig. 1 of \citet{2005AJ....130.1171Y} illustrates the positions of the components A-E \citep[according to the identifications of][]{1983A&AS...51..143G}. Component A \citep[O7.5V,][]{2013hsa7.conf..672S} is the principal source of UV illumination of the central nebula. Component B is an A2Ia supergiant, of approximately the same brightness as component A. HD\,164492D lies only 2 arcsec west of component C and is classified as a Herbig Be star \citep[LkH$\alpha$ 123;][]{2005AJ....130.1171Y}. Considering the angular proximity of the various components, the risk of confusing one star for another, or of contamination of an observation by the seeing disc or scattered light of another, is potentially serious.

Evidence of a strong magnetic field present in one or more components of the spectroscopic binary HD\,164492C was reported by \citet{2014A&A...564L..10H}. Those authors employed optical circular polarisation spectroscopy obtained with FORS2 (on the VLT) and HARPspol (on the ESO 3.6m telescope at La Silla) to obtain convincing detections of Zeeman signatures in the composite spectral lines of the system. HD\,164492C is an early-type spectroscopic binary system. Hubrig et al. identified the presence of at least two contributions to the line profiles. They noted the presence of a weak He~{\sc ii} $\lambda 4686$ line in "the central, strongest component" and assigned it a spectral type B1 or B1.5 (depending on the He abundance), corresponding to effective temperature ($T_{\rm eff}$) of 24-26 kK. They speculated that the system might have a third component based on the complexity of the line profiles they observed.

The BinaMIcS project \citep{2015IAUS..307..330A,2016A&A...589A..47A} aims to understand the interplay between magnetism and binary tidal interaction, by determining the magnetic properties of massive short-period binary systems, and comparing them to isolated stars. It is in this framework that we performed a spectroscopic and spectropolarimetric study of this newly discovered magnetic binary system.

In this paper, we exploit new and archival spectropolarimetric and spectroscopic observations of HD\,164492C to better understand the structure of the system and the physical characteristics of the components, including their magnetic properties. In Sect.~\ref{obs} we describe the observational material employed in our study. In Sect.~\ref{specrvs} we describe the spectral properties of the system, and the radial velocities of the previously identified components. In Sect.~\ref{compc} we search for, and ultimately identify, the third stellar component of the system. In Sect.~\ref{orbit} we model the orbit of the double-lined spectroscopic binary (SB2). In Sect.~\ref{magnetic} we derived the physical properties of the broad-lined magnetic star (and estimate those of the other components of the system), as well as its magnetic geometry. We also model the variability of the H$\alpha$ emission, and constrain the properties of the star's centrifugal magnetosphere. Finally, in Sect.~\ref{conclusion} we discuss the limitations of our results and their broader implications, along with outstanding questions and issues.

\begin{figure}
\hspace{-0.25cm}\includegraphics[width=7.4cm,angle=-90]{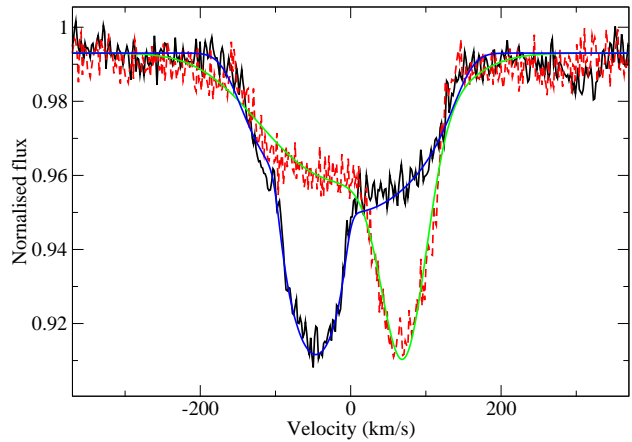}
\caption{Two profiles of the Si~{\sc iii} $\lambda 4552$ line in the ESPaDOnS spectra, along with representative multiline fits for determination of radial velocities. There are two obvious contributions to the profile: a narrow-lined profile (C2) that changes radial velocity significantly between the two spectra, and a broad-lined component (C1) that appears to be stationary. The profile fits - particularly those corresponding mainly to the broad-lined star - show some differences in shape and width due to the variable profile shape. These uncertainties introduce scatter into the radial velocity measurements of that star.}\label{Si3line}

\end{figure}

\section{Observations}\label{obs}

\subsection{ESPaDOnS and HARPSpol spectropolarimetry}

Seventeen high resolution circular polarisation (Stokes $I$ and $V$) spectra were obtained in 2015 with the ESPaDOnS instrument at the Canada-France-Hawaii Telescope (CFHT) in the context of the Binarity and Magnetic Interactions in Stars (BinaMIcS) Large Program. Observations were obtained between June 24 and July 4 2015. In addition, we exploited 6 HARPSpol Stokes $V$ spectra obtained from the European Southern Observatory (ESO) archive. (These spectra were obtained by the BoB Collaboration \citep{2015IAUS..307..342M}, and some were already reported and analysed by \citealt{2014A&A...564L..10H}).  Both instruments are fibre fed, high resolution spectropolarimeters. Each ESPaDOnS polarimetric observation was obtained by combining 4 successive subexposures of 840s, between which the Fresnel rhombs were rotated. The total duration of each ESPaDOnS observation was 3600s. Those data were reduced at the CFHT using the Upena pipeline, and normalized by interactive fitting of polynomial functions to hand-selected continuum regions in each spectral order. The ESPaDOnS data have a resolving power $R=65000$, and span a spectral range from 380 nm to 1~micron. The HARPSpol observations, acquired in the context of observing program 191.D-0255, used a similar observing procedure. The subexposure times were 2700s, and the total observation time per spectropolarimetric sequence was 10800s. Observations were reduced using the the {\sc reduce}-based procedure described by \citet{2011A&A...536L...6A}. The HARPSpol observations have a resolving power of about 105000, and span the wavelength range 378 nm to 691 nm with a gap between 526 and 534 nm. The general details of the HARPSpol observations are described by \citet{2014A&A...564L..10H}. The log of ESPaDOnS and HARPSpol observations is reported in Table~\ref{specpol_table}.

\begin{table*}
\begin{centering}
\caption{Log of spectropolarimetric observations. S/N indicates the peak signal-to-noise ratio per reduced pixel in the diagnostic null spectrum. Rotational phases are computed from Eq. 2. Detection flags refer to the detection status from the False Alarm Probabilities (FAPs), either a definite detection ({\bf DD}: ${\rm FAP} < 10^{-5}$), marginal detection (MD: $10^{-5} < {\rm FAP} < 10^{-3}$), or non-detection (ND: ${\rm FAP} > 10^{-3}$).}\label{specpol_table}
\begin{tabular}{ccrcrcrc}
\hline
UT date & HJD & S/N & Rotational & $\langle B_{\rm z}\rangle$ & Detection & $\langle N_{\rm z}\rangle$ & Detection \\
 &-2400000 & & phase  & (G) & Flag & (G) & Flag \\
\hline
\multicolumn{8}{c}{ESPaDOnS}
\\
2015-06-24 & 57197.93481 &  545 & 1.00000 & 1838$\pm$144 & {\bf DD} & 5$\pm$144 & ND \\
2015-06-24 & 57197.97860 &  538 & 0.03197 & 1767$\pm$147 & {\bf DD} & -126$\pm$147 & ND \\
2015-06-24 & 57198.02331 &  522 & 0.06461 & 1896$\pm$154 & {\bf DD} & 180$\pm$154 & ND \\
2015-06-28 & 57201.93525 &  531 & 0.92032 & 1860$\pm$150 & {\bf DD} & 18$\pm$150 & ND \\
2015-06-28 & 57201.97607 &  531 & 0.95012 & 1796$\pm$150 & {\bf DD} & -147$\pm$150 & ND \\
2015-06-29 & 57202.80395 &  497 & 0.55447 & -381$\pm$149 & ND & 116$\pm$149 & ND \\
2015-06-29 & 57202.84476 &  489 & 0.58426 & -227$\pm$154 & ND & -35$\pm$154 & ND \\
2015-06-30 & 57203.88685 &  513 & 0.34499 & 174$\pm$152 & {\bf DD} & -101$\pm$152 & ND \\
2015-06-30 & 57203.92767 &  490 & 0.37479 & -106$\pm$162 & {\bf DD} & -123$\pm$162 & ND \\
2015-07-01 & 57204.82306 &  470 & 0.02842 & 1723$\pm$189 & {\bf DD} & -62$\pm$189 & ND \\
2015-07-01 & 57204.86385 &  483 & 0.05820 & 1968$\pm$181 & {\bf DD} & -134$\pm$181 & ND \\
2015-07-02 & 57205.82667 &  546 & 0.76106 & 1193$\pm$146 & {\bf DD} & 73$\pm$146 & ND \\
2015-07-02 & 57205.86748 &  553 & 0.79085 & 1237$\pm$144 & {\bf DD} & 10$\pm$144 & ND \\
2015-07-03 & 57206.95937 &  448 & 0.58793 & -218$\pm$175 & ND & -73$\pm$175 & ND \\
2015-07-03 & 57207.00017 &  421 & 0.61771 & 51$\pm$196 & ND & -12$\pm$196 & ND \\
2015-07-04 & 57207.86580 &  404 & 0.24962 & 973$\pm$195 & {\bf DD} & -181$\pm$195 & ND \\
2015-07-04 & 57207.90661 & 1457 & 0.27941 & 871$\pm$238 & {\bf DD} & 275$\pm$238 & ND \\
\hline
\\
\multicolumn{8}{c}{HARPSpol}
\\
2014-04-22 & 56770.79323 &  205 & 0.18665 & 1142$\pm$160 & {\bf DD} & 239$\pm$160 & ND \\
2014-04-23 & 56771.82098 &  225 & 0.93690 & 2067$\pm$150 & {\bf DD} & 253$\pm$149 & ND \\
2015-03-09 & 57091.84260 &  180 & 0.55269 & -448$\pm$193 & ND & -180$\pm$193 & ND \\
2015-03-10 & 57092.82821 &  187 & 0.27218 & 882$\pm$183 & {\bf DD} & 214$\pm$183 & ND \\
2015-03-11 & 57093.84781 &  208 & 0.01649 & 1881$\pm$164 & {\bf DD} & -63$\pm$164 & ND \\
2015-03-12 & 57094.81930 &  160 & 0.72568 & 937$\pm$214 & ND & 0$\pm$214 & ND \\
\hline\hline
\end{tabular}
\end{centering}
\end{table*}

\subsection{UVES and FEROS spectroscopy}

UVES \citep{2000SPIE.4008..534D} spectra of HD 164492C were obtained on 15 nights in 2014, under ESO program 093.D-0267. In addition, 4 FEROS \citep{1999Msngr..95....8K} spectra of the system were obtained in 2013, under ESO program 091.C-0713. All raw data were downloaded from the ESO archive. FEROS spectra were reduced with the ESO-MIDAS based data reduction system available on the web\footnote{www.eso.org/sci/facilities/lasilla/instruments/feros/tools/DRS.html}. UVES spectra were reduced with the the ESO-reflex-based UVES pipeline \citep[see][]{2013A&A...559A..96F}. The merged spectra were normalized with a spline interpolation based on hand-selected continuum regions. The FEROS data have a resolving power $R=48000$, and range in wavelength from about 380~nm to 920~nm. The UVES data were obtained with wavelength settings of 390/760 using a slit width of 0.4" in the blue arm and 0.3" in the red arm. This yielded $R\sim 80000$ for the blue arm and $R\sim 90000$ for the red arm, with respective wavelength ranges of $326-445$~nm and $560-946$~nm.

The log of the FEROS and UVES spectra that were used in this study is provided in Table~\ref{uves_table}.

\begin{table}
\begin{centering}
\begin{tabular}{cccc}
\hline
Instrument & UT date & JD & S/N \\
& &-2400000 &    \\
\hline
UVES   &    2014-04-25    &    56772.770    & 280     \\
UVES   &    2014-04-30    &    56777.808    & 252     \\
UVES   &    2014-05-02    &    56779.895    & 162     \\
UVES   &    2014-05-09    &    56786.881    & 410     \\
UVES   &    2014-06-29    &    56837.604    & 265     \\
UVES   &    2014-07-31    &    56869.507    & 361     \\
UVES   &    2014-08-10    &    56879.697    & 301     \\
UVES   &    2014-08-13    &    56883.487    & 317     \\
UVES   &    2014-08-17    &    56887.493    & 277     \\
UVES   &    2014-08-25    &    56894.626    & 269     \\
UVES   &    2014-08-29    &    56898.568    &   279  \\
UVES   &    2014-09-19    &    56919.560    & 279     \\
UVES   &    2014-09-23    &    56923.511    &  325   \\
UVES   &    2014-09-26    &    56926.538    & 341     \\
UVES   &    2014-09-28    &    56929.499    &  376   \\
FEROS    &   2013-08-18    &    56522.684  &   160   \\
FEROS    &   2013-08-19    &    56523.703  &    125  \\
FEROS    &   2013-08-20    &    56524.622   &      115\\
FEROS    &   2013-08-21    &    56525.662  &    105  \\\hline\hline
\end{tabular}
\caption{Table of UVES and FEROS spectra. UVES signal-to-noise ratio (S/N) is the mean S/N per pixel measured in blue order 25. FEROS S/N was measured in the continuum near the blue Si~{\sc iii} lines.}\label{uves_table}
\end{centering}
\end{table}

\section{Spectrum and radial velocities}\label{specrvs}

\begin{figure*}
\begin{tabular}{ccc}
\includegraphics[width=5.5cm]{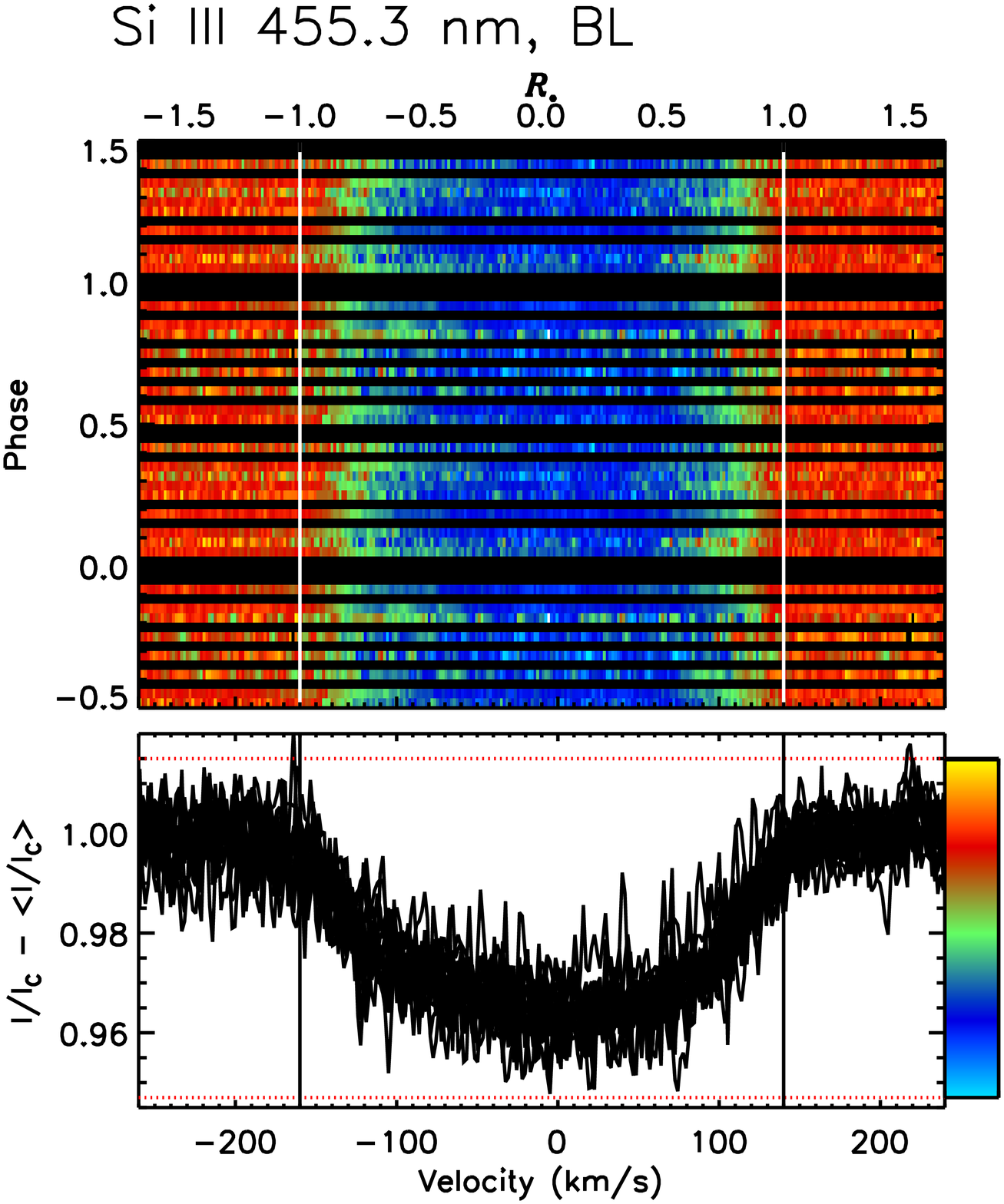} &
\includegraphics[width=5.5cm]{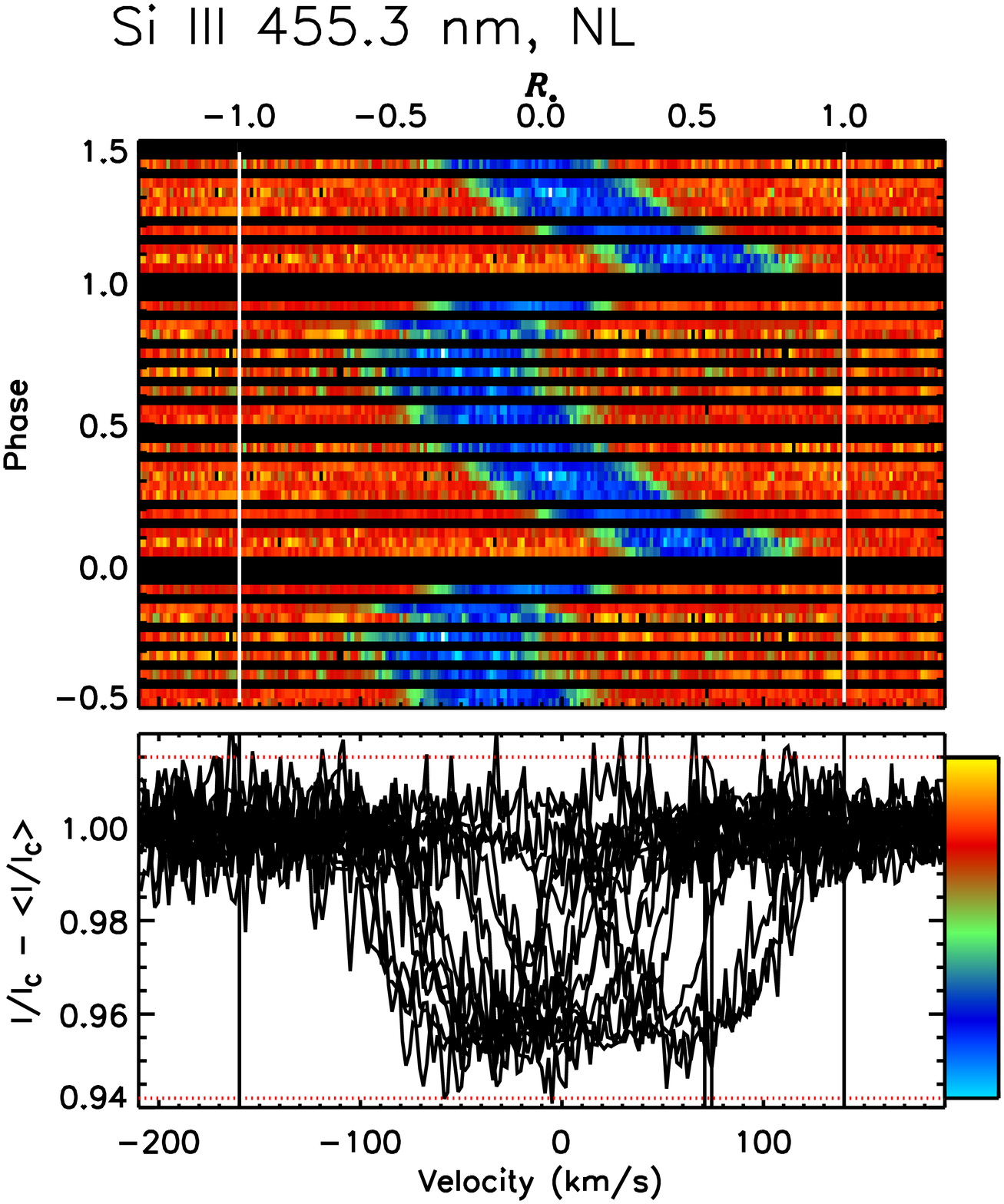} &
\includegraphics[width=5.5cm]{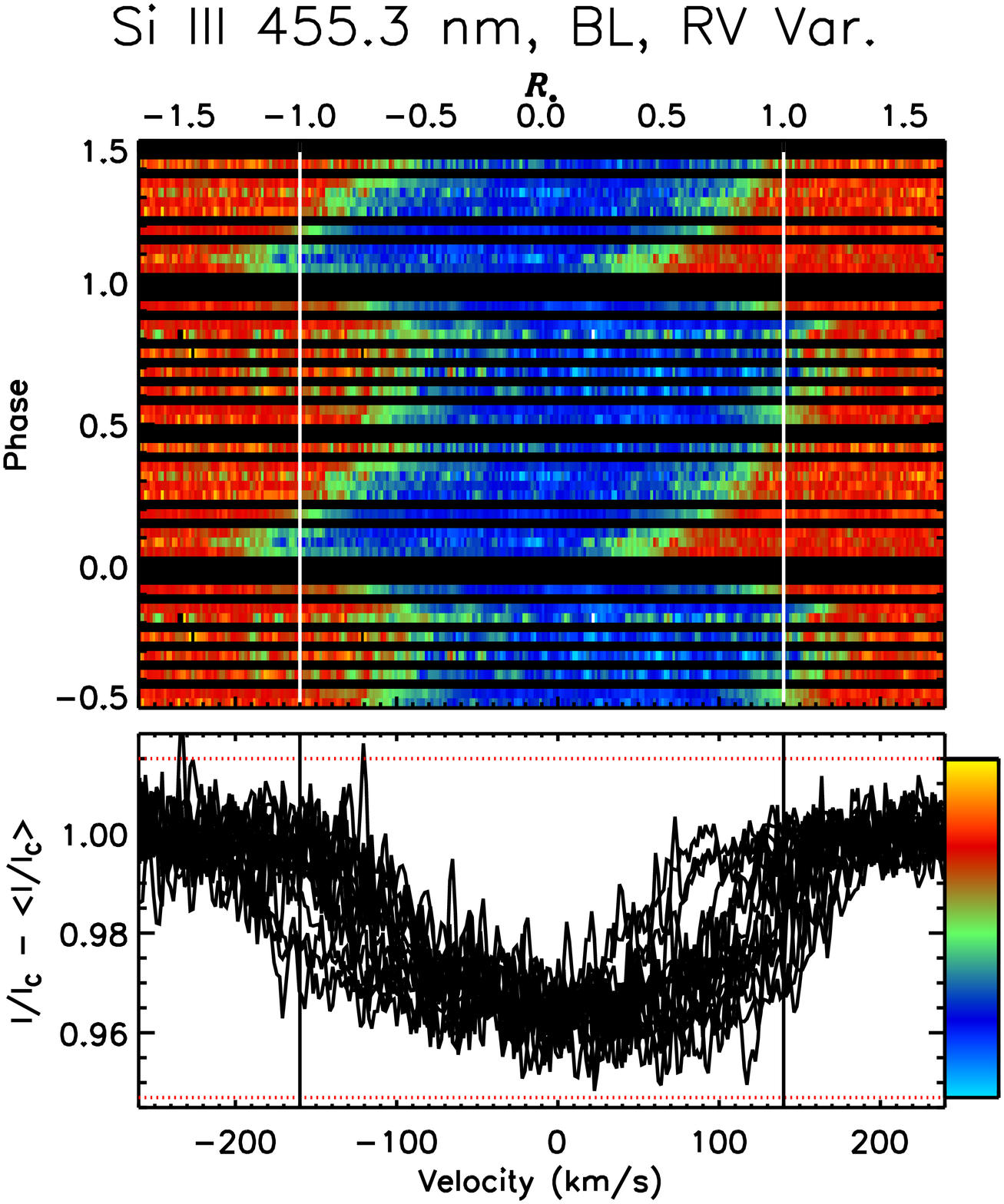} \\
\end{tabular}
\caption{Dynamic spectra of the disentangled Si~{\sc iii} $\lambda 4552$ profiles of the C1 (BL='Broad lined') and C2 (NL='Narrow-lined') components (left and middle, respectively), phased according to the 12.53d period. The right panel shows the dynamic spectrum of the C1 component as it would appear with an artificial RV variation equal to $-0.8\times$ the RV of the C2 component, as would be expected if the C1 component is slightly more massive. Such a RV variation would have been detectable in our data.}\label{disentangle_dyn}
\end{figure*}

The spectrum of the system is clearly composite, showing obvious profiles of at least two stars. Fig.~\ref{Si3line} shows the profile of the Si~{\sc iii} $\lambda 4552$ line in two of the ESPaDOnS spectra obtained on different dates. Two profile contributions are visible: one relatively broad and one significantly narrower. For clarity, we henceforth will refer to the broad-lined component as HD\,164492C1, and the narrow-lined component as HD\,164492C2. As is evident from the figure, component C2 exhibits significant bulk radial velocity (RV) variability, consistent with orbital motion. However, visual examination of all spectra suggests that the profile of C1 is stationary, and exhibits no comparable RV changes.

Both stars have well-developed Si~{\sc iii} profiles. In agreement with \citet{2014A&A...564L..10H}, we also note the presence of a weak feature at the location of the He~{\sc ii} $\lambda 4686$ line. This feature does not move in RV in our spectra. Based on the RV behaviour of the Si~{\sc iii} profiles, we associate the weak He~{\sc ii} line with the C1 component. (It is unclear to us, in the discussion by Hubrig et al., to which component they attribute this line.) While C2 may make a marginal contribution to this feature, it does not appear to contribute significantly, in the sense that we observe no obvious variable contribution to it.

The general characteristics of the broad-lined and narrow-lined spectral contributions are consistent with those described by Hubrig et al. In particular, both stars appear to be of early B spectral type. The C2 component may be marginally cooler and fainter. The similar equivalent widths of the Si~{\sc iii} and C~{\sc ii} lines of the two stars are consistent with this conclusion, and furthermore suggest that the apparent magnitudes of the two stars in the $V$ band are not very different.

We first measured the RVs of the two stars by fitting the profiles of relatively strong, effectively unblended spectral lines showing clear contributions from both components. We used the Si~{\sc iii} $\lambda 4552, 4567$ lines, the C~{\sc ii}~$\lambda 4267$ line and the He~{\sc i} $\lambda 6678$ lines. The UVES spectra did not contain the Si~{\sc iii} lines, so only the C~{\sc ii} and He~{\sc i} lines were measured in those spectra. We employed the {\sc idl} fitting tool described by Grunhut et al. (2016, submitted) to simultaneously fit synthetic profiles (including rotational and turbulent broadening) to the observed profiles. An example of such fits is shown in Fig~\ref{Si3line}. We estimated RV uncertainties using the dispersion of measurements of the different lines in spectra acquired on the same night. This amounted to about 2-3 km/s for the narrow-lined component, and 5-10 km/s for the broad-lined component. 

The C2 component is observed to vary significantly and coherently in velocity by over 130 km/s. (This will be discussed further in Sect.~\ref{orbit}.) A period search of these RV measurements using a Lomb-Scargle approach yields a single strong peak in the periodogram near 12.53d.

On the other hand, and as expected from Fig.~\ref{Si3line}, no coherent variation of the RV of the broad-lined component is detected. However, precise measurement of the broad-lined star's RV is challenging: the narrow-lined star often hides one or the other wing of the broad line, leading to best-fit model profiles corresponding to C1 that that exhibit important differences in width and position. Ultimately, these ambiguities introduce significant scatter into the RV measurements of the broad-lined star. The dispersion of the measurements was about 14 km/s. 

To investigate whether a coherent variation of the broad spectral line could be hiding in these uncertainties, we performed iterative line profile disentangling using the method introduced by \cite{gonzalez_levato_2006}. Disentangling begins with the measured RVs and approximate models of the line profiles (\vsini~and $v_{\rm mac}$) of the components. Model line profiles of individual observations are obtained by shifting the model profiles of the component stars to the measured RVs, and then combining them with an assumed EW ratio, with the composite profile normalized to the observed EW. The normalized model line profiles of the first component are then subtracted from the observations, the residual flux of each observation is shifted by subtracting the RV of the second component, and a mean line profile is obtained by adding the shifted residual profiles. This procedure provides an approximate line profile for the second component. The same process is then repeated with model profiles of the second component in order to obtain an approximation for the first component's profile. The mean disentangled line profiles obtained from the previous iteration are then used as input for the next iteration, and the process is repeated until the $\chi^2$ of the fit to the individual observations ceases to change significantly from one iteration to the next.

Disentangling was performed both by adhering to the measured RVs, and by introducing RV refinement by re-measuring the RVs at each step using the centre-of-gravity of the residual line profiles. In our experience, the latter method can be successful only if the initial model and RVs provide a reasonable approximation to the shape and behaviour of the line profiles: if not, the solution will fail to converge. Our initial model used 140 \kms~of \vsini~for the broad-lined component, \vsini~$=50$~\kms~for the narrow-lined component, 20 \kms~of radial-tangential macroturbulence for both, and EW fractions of 0.7 and 0.3 for the two components, respectively. Re-measuring the RV did not result in the detection of any significant RV variability in the broad-lined component, with a final standard deviation of the output RVs of 4.7 \kms, as compared to 41 \kms~for the narrow-lined component. The disentangled profiles of the broad and narrow-lined components of the Si~{\sc iii}~$\lambda 4552$ line are shown as dynamic spectra phased with the 12.53d period in the left and middle panels of Fig.~\ref{disentangle_dyn}. 

As an additional check that no comparable systematic RV variation of the broad-lined star could be present, we added an artificial RV variation to the disentangled profiles of C1 equal to $-0.8\times$ the RV of the narrow-lined component, as would be expected if the two stars orbit a common centre-of-mass, with the broad-lined component being slightly more massive (e.g. Sect.~\ref{subsec_bzmodel}). The resulting dynamic spectrum is shown in the right panel of Fig.\ \ref{disentangle_dyn}, and demonstrates that such RV variability would have been detected if it were present. Furthermore, to verify that our initial assumption of zero RV variation of the broad-lined star was not affecting the disentangling process, we also attempted to disentangle the profiles using this same artificially introduced RV variation. No acceptable convergence could be obtained, with the resulting $\chi^2/\nu=3.1$, significantly in excess of the values obtained using the measured RVs (1.27), a constant RV (1.26), or recalculated RVs (1.26).

The adopted RVs of the C1 and C2 components are presented in Table~\ref{RV_table}. In this Table, we report the RVs obtained from the He~{\sc i} $\lambda 6678$ line (which could be obtained for both components from all spectra), refined to provide the best fit to all profiles following disentangling. Two measurements of the RV of C1 (at phase $\sim 0.03$) may stand out visually from the remainder of the data for that star. The measurements (obtained on the same night) depart form the mean by roughly $2\sigma$. We have investigated the significance of these results (by re-initialising the disentangling using slightly different model parameters, and by extending the procedure over additional iterations), and conclude that these departures from the mean are not significant.

From our disentangling, we ultimately conclude that there is no evidence for significant RV variations of the broad-lined star, with a standard deviation of about 5 km/s. A Lomb-Scargle period search of these measurements yields no significant signal, in particular near the 12.53d period detected in the narrow-lined star's measurements. Were the broad-lined and narrow-lined stars in a physical binary system, the measured RVs would imply a mass ratio of about 13:1, which is strongly inconsistent with the similar spectral types and apparent magnitudes of the stars. We therefore conclude that the broad-lined star is not the companion at the origin of the narrow-lined star's binary motion. As a consequence, in the following Section we search the spectra for evidence of a third component.

\begin{table}
\begin{centering}
\begin{tabular}{ccrrr|ccc|cc}
\hline
HJD    &    Orbital    &    C1    &    C2    &    C3    \\
-2400000    &    phase    &    (km/s)    &    (km/s)    &    (km/s)    \\
\hline
\multicolumn{5}{c}{\bf ESPaDOnS}            \\
57197.935    &    0.557   &  -0.5        &    -30.8    &    46.8    \\
57197.979    &    0.561    &  -0.3        &    -31.2    &    52.0    \\
57198.023    &    0.564    &  -0.2      &    -30.3    &    51.7    \\
57201.935    &    0.877    &   0.6      &    -52.2    &    89.7    \\
57201.976    &    0.880    &   1.1        &    -52.7    &    88.7    \\
57202.804    &    0.946    &   1.0        &    -21    &        \\
57202.845    &    0.949    &   0.1      &    -16.3    &        \\
57203.887    &    0.032    &   0.3       &    75.5    &    -184.0    \\
57203.928    &    0.036    &   0.0       &    78.1    &    -187.2    \\
57204.818    &    0.107    &  -0.0       &    67    &    -154.4    \\
57204.864    &    0.110    &   0.4       &    64.2    &    -151.5    \\
57205.827    &    0.187    &  -0.0       &    35.1    &    -102.5    \\
57205.867    &    0.190    &  -1.0        &    33.1    &    -103.4    \\
57206.959    &    0.278    &   1.2       &    5.3    &        \\
57207.000    &    0.281    &  -0.3       &    4.5    &        \\
57207.866    &    0.350    &  -1.2        &    -6.2    &        \\
57207.907    &    0.353    &  -1.0        &    -6.6    &        \\
\multicolumn{5}{c}{\bf UVES}              \\
56772.771    &    0.629    &  -0.3      &    -41.8    &    71.0    \\
56777.808    &    0.031    &  -1.5       &    79.5    &    -176.8    \\
56779.895    &    0.197    &  -1.8        &    28.2    &    -82.3    \\
56786.882    &    0.755    &  -0.6        &    -54    &    75.7    \\
56837.604    &    0.803    &  -0.3        &    -56.6    &        \\
56869.508    &    0.348    &  -3.4        &    -7.6    &        \\
56879.697    &    0.162    &  -2.0        &    40.3    &        \\
56883.487    &    0.464    &  -3.3      &    -17.6    &        \\
56887.493    &    0.784    &  -1.0        &    -55.3    &    96.5    \\
56894.626    &    0.353    &  -0.9        &    -5.3    &        \\
56898.568    &    0.668    &   1.4        &    -45.8    &    83.8    \\
56919.561    &    0.343    &  -3.6        &    -5.4    &        \\
56923.511    &    0.658    &  -0.3        &    -42.4    &    59.182    \\
56926.538    &    0.900    &  -0.9        &    -44.1    &    69.7    \\
\multicolumn{5}{c}{\bf HARPSpol}       \\
56770.793    &    0.471    &  -2.1        &    -20.7    &        \\
56771.821    &    0.553    &  -0.6        &    -32.3    &    56.0    \\
57091.843    &    0.091    &  -0.1     &    67.6    &    -159.0    \\
57092.828    &    0.170    &  -2.7        &    39    &    -108.6    \\
57093.848    &    0.251    &  -0.1        &    14.2    &        \\
57094.819    &    0.329    &  -0.3      &    -3.5    &        \\
\multicolumn{5}{c}{\bf FEROS}              \\
56522.684    &    0.671    &   0.2        &    -49.9    &    74.2    \\
56523.704    &    0.753    &   0.0        &    -54.5    &    76.8    \\
56524.622    &    0.826    &   0.5        &    -56    &    85.8    \\
56525.662    &    0.909    &  -1.6        &    -36.5    &        \\
\hline\hline
\end{tabular}
\caption{Table of radial velocities. The orbital phase is computed using the period and zero-point JD reported in Table~\ref{orbital_params}. At some phases, the weak profile of the C3 component was sufficiently blended with those of the other stars that its RV could not be reliably measured.}\label{RV_table}

\end{centering}
\end{table}

\section{Search for an additional spectroscopic component}\label{compc}

Considering the large RV variation exhibited by the narrow-lined star, we searched our spectra for potentially subtle contributions from an additional component in the system. 

We used the collection of reduced spectra to search for the presence of a third component. No clear evidence was found when visually examining the majority of the observed spectral lines. This included all of the strongest lines for which both the C1 and C2 components could be easily discerned (e.g. the Balmer, He~{\sc i}, and Si~{\sc iii} lines). We therefore attempted to model the observed spectra assuming the presence of only the two obvious components.

This was carried out using the grid of non-local thermodynamic equilibrium (NLTE) {\sc tlusty} {\sc bstar2006} spectral models \citep{Lanz2007}. The grid consists of models having $T_{\rm eff}$ ranging from $15-30\,{\rm kK}$ in increments of $1\,{\rm kK}$ and surface gravity ($\log{g}$) ranging from a minimum of $2.0$ to a maximum of $4.75\,{\rm (cgs)}$ in increments of $0.25\,{\rm (cgs)}$.

We adopted initial effective temperatures of the two components of $25\,{\rm kK}$ based on the previous conclusion that both stars are of early B spectral type. The $\log{g}$ values of both components were fixed at $4.0\,{\rm (cgs)}$ and a solar metallicity was assumed. No macroturbulence or microturbulence was included in the models. Approximate $v\sin{i}$ values of $140$ and $50\,$\kms were found to be appropriate based on comparisons with, in particular, the Si~{\sc iii} lines. All of the synthetic spectra were convolved with a Gaussian function appropriate to the instrumental broadening of each (ESPaDOnS, HARPSpol, UVES or FEROS) spectrum.

The total flux was initially calculated assuming a luminosity ratio of unity between the two components. Comparisons between this total model flux and the observed Balmer lines -- specifically H$\beta$, H$\gamma$, and H$\delta$ -- were then used to better constrain each component's adopted $T_{\rm eff}$; H$\alpha$ was not used because of the clear emission present in the wings (this is discussed further later in the paper). We note that the cores of H$\alpha$, H$\beta$, and H$\gamma$ all exhibit narrow emission which is assumed to be associated with hot gas within the Trifid nebula. Ultimately, we found that using $T_{\rm eff}=26\,{\rm kK}$ and $T_{\rm eff}=24\,{\rm kK}$ for the broad-line and narrow-line components, respectively, yielded the best-fitting solution as determined by eye. 

\begin{figure}
\includegraphics[width=8cm]{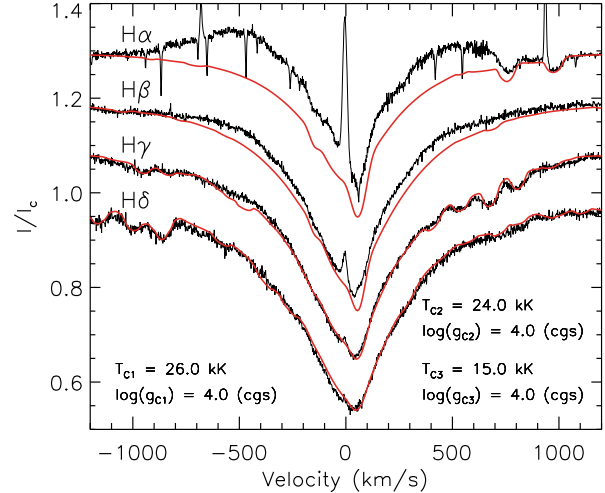}
\caption{Comparisons between the observed spectrum (black) and the total {\sc tlusty} model (red) including all three components. Four H Balmer lines are shown; from top to bottom: H$\alpha$, H$\beta$, H$\gamma$, and H$\delta$.}\label{model_spec_Balmer}
\end{figure}

We found that this model adequately reproduced most of the strongest lines in the spectrum. The model was found to be consistent with the inner wings (i.e. from $\approx100-500\,$\kms) of H$\beta$, H$\gamma$, and H$\delta$ (Fig.~\ref{model_spec_Balmer}).  Many of the He~{\sc i} lines (e.g. He~{\sc i} $\lambda 4144$, He~{\sc i} $\lambda 4388$, and He~{\sc i} $\lambda 4471$) showed broad wings that were not reproduced by the model. We suspect that this is a consequence of a peculiar abundance distribution of He, implying the presence of large-scale overabundance patches and/or strong vertical stratificaton of this element. This will be discussed further in Sect.~\ref{conclusion}.

\begin{figure}
\includegraphics[width=8cm]{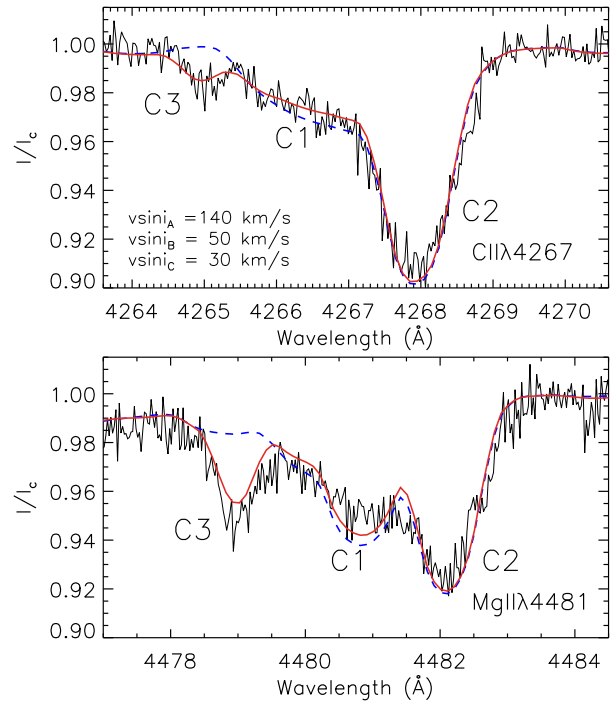}
\caption{\emph{Top:} C~{\sc ii}$\lambda$4267 line used to determine the 
luminosity ratios between the three components. \emph{Bottom:} Mg~{\sc ii}$\lambda$4481 line. The black curves correspond to the observed spectrum, the solid red curves correspond to the total {\sc tlusty} model including all three components, and the dashed blue curves show the total model spectrum when only the two brightest components are considered. The components, labeled as C1, C2, and C3, are identified in the text.}\label{model_spec_CIIMgII}
\end{figure}

\begin{figure}
\includegraphics[width=8cm]{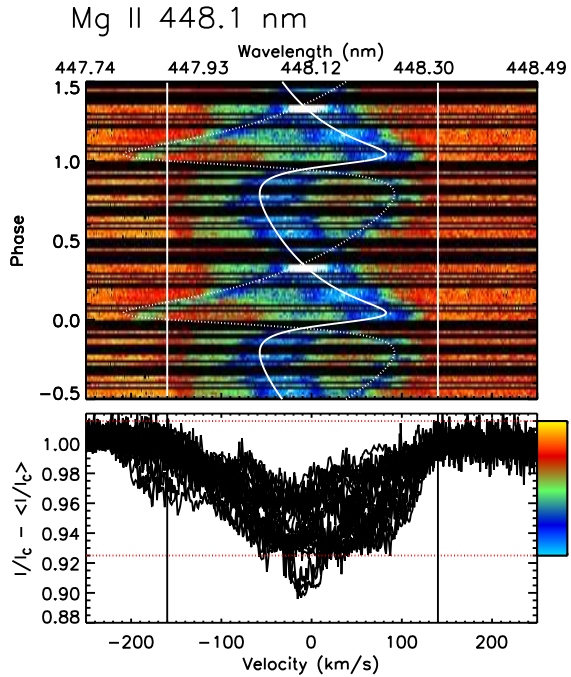}
\caption{Dynamic spectrum of the Mg~{\sc ii} $\lambda 4481$ line showing the coherent RV variations of the C2 and C3 components according to the 12.53d (binary) period. The solid and dotted lines are the respective variations predicted by the orbital model discussed in Sect.~\ref{orbit}. }\label{4481dyn}
\end{figure}

Most remarkably, both C~{\sc ii} $\lambda 4267$ and Mg~{\sc ii} $\lambda 4481$ were observed to exhibit an additional weak and isolated absorption component in their profiles (see Fig. \ref{model_spec_CIIMgII}). This persistent feature was detected at RVs ranging from $\approx-140$ to $+90\,$~\kms -- located approximately in anti-phase with respect to the RV of the C2 component. The dynamic spectrum of the Mg~{\sc ii} $\lambda 4481$ line, phased according to the 12.53d period, is shown in Fig.~\ref{4481dyn}. This clearly shows the phased orbital motion of the C2 component (overlaid with the orbital model (solid line) discussed later in Sect.~\ref{orbit}). The dynamic spectrum also shows the more subtle motion of the newly-detected spectral feature, which exhibits a fainter trace of comparable shape, with a much larger ($\sim 2\times$) RV amplitude (this component's motion is also overlaid with the associated orbital model discussed in Sect.~\ref{orbit}). These properties strongly support the identification of this feature as a tertiary component of the system, which we label HD\,164492C3.

In order to include this third component in the model spectrum, we needed to determine both its $T_{\rm eff}$ and $v\sin{i}$ values. $T_{\rm eff}$ was estimated by first estimating its mass with respect to the $24\,{\rm kK}$ C2 component, using the ratio of RV amplitudes $K_3/K_2\approx 2$ (as estimated from Fig.~\ref{4481dyn}). Under the assumption that both components lie on the main sequence, we found that the mass ratio, i.e. $M_2/M_3=K_3/K_2$, implies a cooler temperature of $\approx15\,{\rm kK}$. A $v\sin{i}$ of $30\,$\kms was estimated by eye based on comparisons between observed and computed profiles of the observed C~{\sc ii} $\lambda 4267$ and Mg~{\sc ii} $\lambda 4481$ lines. This cooler temperature, combined with the associated lower luminosity, explains why this component is not detected in profiles of most strong lines in the spectrum of HD\,164492C.

Calculation of the total model spectrum requires not only $T_{\rm eff}$ and $v\sin{i}$ of each of the three components, but also each component's fractional contribution to the total flux (i.e. $F_{\rm tot}=\sum L_iF_i/L_{\rm tot}$ where $F_i$ corresponds to the $i^{\rm th}$ component's flux, $L_i$ is its luminosity, and $F_{\rm tot}$ and $L_{\rm tot}$ are the system's total flux and luminosity). This was estimated using a procedure developed for analysis of the structurally similar triple system HD~35502 \citep{Sikora2016}. This involved allowing the luminosity ratios of the secondary and tertiary components with respect to the primary ($L_1/L_2$ and $L_2/L_3$, respectively) to be fit based on comparisons with the observed C~{\sc ii} $\lambda 4267$ line. Using only those observations for which the three components were sufficiently separated in velocity, our analysis implies an acceptable ratio $L_1/L_2$ of $1.1\pm 0.1$, and a ratio $L_2/L_3$ of $6.5\pm 0.5$. 

The final model is shown in Figs.~\ref{model_spec_Balmer} and \ref{model_spec_CIIMgII}. Fig.~\ref{model_spec_CIIMgII} shows two models: one in which all three components are considered and one in which only the (previously identified) C1 and C2 components are shown.

\section{Orbit of the binary system}\label{orbit}

The qualitative characteristics of the cooler C3 component detected in the high resolution spectra are fully consistent with the properties expected based on the orbital motion of the C2 component (and the lack of motion of C1). We used the spectral fits to estimate the radial velocity of the third component in each of our spectra. These values - reported in Table \ref{RV_table} - are uncertain at roughly 5 km/s due to the weakness of the spectral contribution of this component and blending from the other components. 

\begin{figure}
\hspace{-0.25cm}\includegraphics[width=8.9cm]{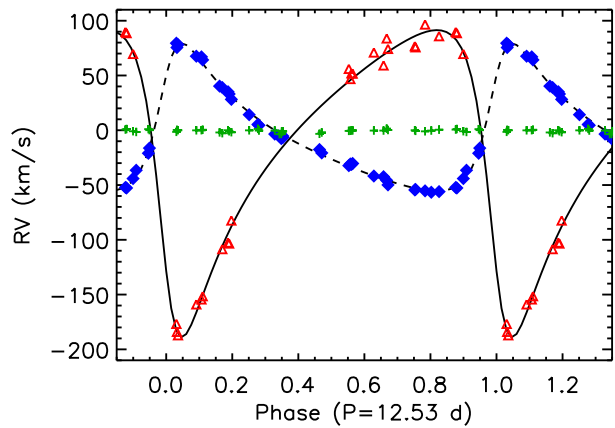}
\caption{Orbital model with RVs, SB2 solution. Blue circles: C2 component. Red triangles: C3 component. Solid and dashed curves: predictions of the respective orbital models. We also show the nonvariable RVs of the C1 component (green plus symbols) for comparison.}\label{RV_2comp}
\end{figure}

We used the IDL orbital fitting code {\sc Xorbit} to model the orbit. This code determines the best-fitting orbital period $P_{\rm orb}$, time of periastron passage ($T$), eccentricity ($e$), longitude of the periastron ($\omega$), semi-amplitudes of each component's radial velocities ($K_1$ and $K_2$), and
the radial velocity of the center of mass ($\gamma$) (Tokovinin 1992), performing least-squares fits to the measured radial velocities. We used as constraints the RVs of both the C2 and C3 components (Table~\ref{RV_table}). As illustrated in Fig.~\ref{RV_2comp}, the RVs of both stars can be reproduced using a single orbital model with $P_{\rm orb}=12.5351\pm 0.0007$~d and $e=0.515\pm 0.006$. This period is in good agreement with the approximate orbital period derived earlier from the narrow-lined star's RVs alone. The RV semi-amplitude of the C3 component ($K_3=140\pm 2$~km/s) implies a mass ratio $M_2/M_3=2.08\pm 0.06$. The orbital parameters derived from the SB2 solution are summarised in Table~\ref{orbital_params}.

\begin{table}
\begin{centering}
\begin{tabular}{lrr}
\hline
Quantity & Value & Uncertainty \\
\hline
Period (d) & 12.5351 & 0.0007 \\
JD0 (d) & 2457190.98 & 0.03\\
$K_2$ (km/s) & 67.4 & 0.8\\
$K_3$ (km/s) & 140 & 2\\
$V_0$ (km/s) & -7.8 & 0.4\\
RMS$_2$ (km/s) & 3.4\\
RMS$_3$ (km/s) & 6.6 \\
$e$ & 0.515 & 0.006\\
$\omega$ ($\degr$) & -55 & 1\\
$M_2/M_3$ & 2.08& 0.06 \\
$M_2\sin^3 i$ ($M_\odot$) & 4.9& 0.3 \\
$M_3\sin^3 i$ ($M_\odot$)& 2.4 & 0.1 \\
$a\sin i$ (AU) & 0.205  & 0.002\\
\hline\hline
\end{tabular}
\caption{Final orbital solution for the 12.53d binary: results of simultaneous modeling of the RV variations of the C2 and C3 components.}\label{orbital_params}
\end{centering}
\end{table}

\section{Properties of the broad-lined magnetic star}\label{magnetic}

\subsection{Least-Squares Deconvolution, magnetic measurements and their period analysis}\label{LSD}

Visual examination of the Si~{\sc iii} profiles shows a broad Stokes $V$ signature that spans the entire width of the composite profile, and that shows no apparent variation in radial velocity. It therefore appears that the magnetic signature is associated, at least principally, with the broad-lined C1 component.

Least-Squares Deconvolution \citep[LSD;][]{1997MNRAS.291..658D} was applied to the ESPaDOnS and HARPSpol spectra, using the {\sc ilsd} program \citep{2010A&A...524A...5K} and a line mask with 571 spectral lines obtained using an {\sc Extract Stellar} request with $T_{\rm eff}=24$~kK and a depth threshold of 0.1 from the Vienna Atomic Line Database (VALD3, \citealt{1995A&AS..112..525P,ryabchikova1997,kupka1999,kupka2000}). The line mask was adjusted to best reproduce the He and metallic line profiles present in the observed spectra using the cleaning/tweaking procedure \citep[e.g.][Grunhut et al., submitted]{2012PhDT.......265G}, first by cleaning the mask of any lines strongly blended with Balmer lines, telluric features, or instrumental ripples, and second by adjusting the depths of the remaining 367 lines so as to match as closely as possible the observations. The LSD profiles were extracted using a 5.4~\kms~pixel width (i.e.\ $3\times$ the 1.8~\kms~ESPaDOnS pixel width), in order to increase the Stokes $V$ S/N. 

\begin{figure}
\includegraphics[width=8.2cm]{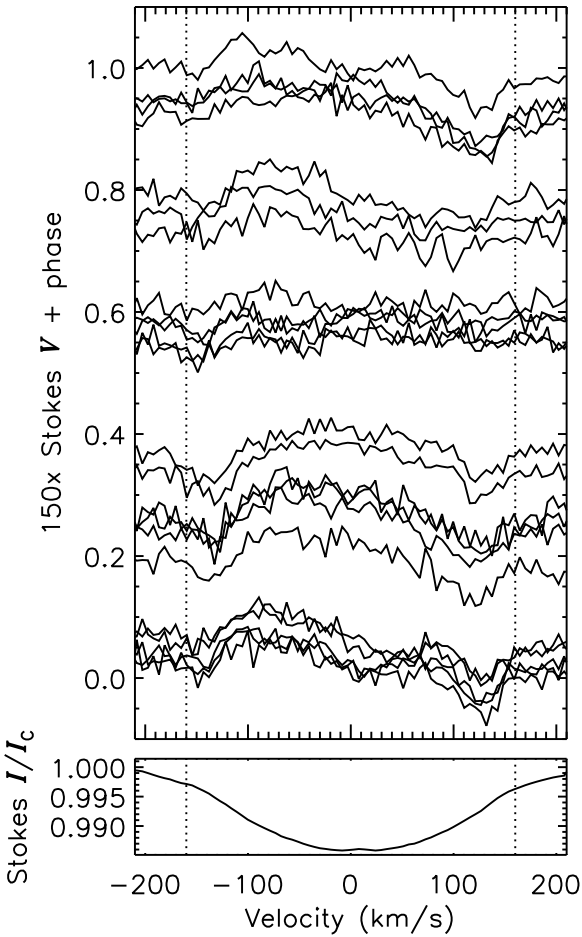}
\caption{LSD Stokes $V$ profiles obtained from a mask containing metallic + He lines, arranged in order of rotational phase. The bottom panel shows the mean Stokes $I$ profile obtained from the disentangled LSD Stokes $I$ profiles of the C1 component. Vertical dotted lines indicate $\pm$\vsini~$+v_{\rm sys}$.}
\label{lsd_stokesv}
\end{figure}

The resulting LSD Stokes $I$ profiles were then disentangled using the same iterative algorithm as used for the Si~{\sc iii} $\lambda 4552$ line. Fig.\ \ref{lsd_stokesv} shows the LSD Stokes $V$ profiles above the mean disentangled Stokes $I$ profile of the C1 component. The Stokes $V$ profiles  - which are not subject to any disentangling - show no evidence of any systematic radial velocity variability, and span the same range in velocity as the line profile of the broad-lined component, strongly supporting our impression that the broad-lined C1 component is the magnetic star. 

To evaluate the significance of detection of Zeeman signatures, we calculated False Alarm Probabilities (FAPs) via comparison of the signal within the Stokes $V$ line profile, as compared to the null hypothesis as evaluated from the noise in the wings and the diagnostic null profile $N$ \citep{1992A&A...265..669D,1997MNRAS.291..658D}. A detection is considered to be definite (DD) if ${\rm FAP} < 10^{-5}$, to be a non-detection (ND) if ${\rm FAP} > 10^{-3}$, and to be a marginal detection (MD) otherwise. Table \ref{specpol_table} provides the detection status for Stokes $V$ and $N$ of each spectrum. All $N$ profiles are NDs, while 17/23 Stokes $V$ profiles yield DDs. 

In the following, based on the evidence above, we assume that the C1 component is the source of the Stokes $V$ signatures shown in Fig.~\ref{lsd_stokesv}, and that contributions from the other components to Stokes $V$ are negligible. We use the LSD Stokes $V$ profiles, along with the disentangled Stokes $I$ profile of the C1 component (i.e. the timeseries of profiles shown in Fig.~\ref{lsd_stokesv}) for our calculations. We emphasise that while an equivalent-width ratio of the lines of the C1 and C2 components was assumed to initialise the disentangling procedure, ultimately our disentangled profiles are not subject to any rescaling or any assumptions about the luminosity ratio of the components. In particular, we preserve the EWs of the profiles as they appear in the composite spectrum, and hence the relationship between the Stokes $I$ and Stoke $V$ profiles.

The mean longitudinal magnetic field \bz\ was measured using \citep{1989FCPh...13..143M,1997MNRAS.291..658D}:

\begin{equation}\label{blong}
\langle B_z  \rangle  = -2.14\times 10^{11}\frac{\int \! vV(v)\mathrm{d}v}{\lambda_0 g_0 c\int \left[I-I(v)\right] \mathrm{d}v},
\end{equation}

\noindent where $v$ is the Doppler velocity relative to the line centre-of-gravity in \kms, and $\lambda_0=500$ nm and $g_0=1.2$ are the normalization values of the wavelength and Land\'e factor used to scale the LSD profiles. Error bars were obtained by propagating the individual pixel error bars through Eqn.\ \ref{blong}. In addition to \bz, \nz~was measured by substituting $N$ for Stokes $V$ in Eqn.\ \ref{blong}. \bz~and \nz~are provided in Table \ref{specpol_table}. The median \bz~measurement is significant at 6.5$\sigma$, while the median \nz~is significant at only 0.5$\sigma$. 

\begin{figure}
\includegraphics[width=8.5cm]{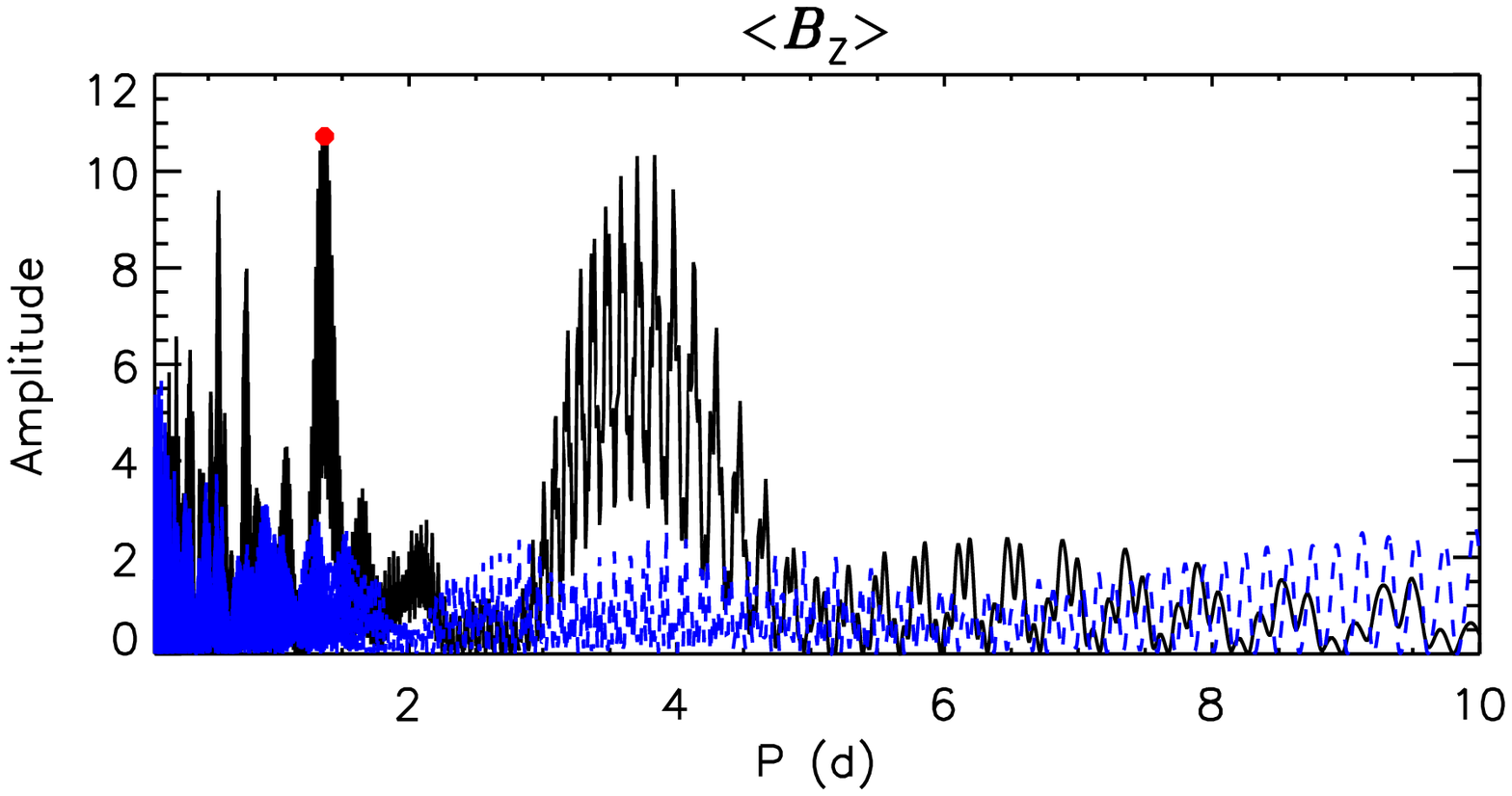}
\includegraphics[width=8.5cm]{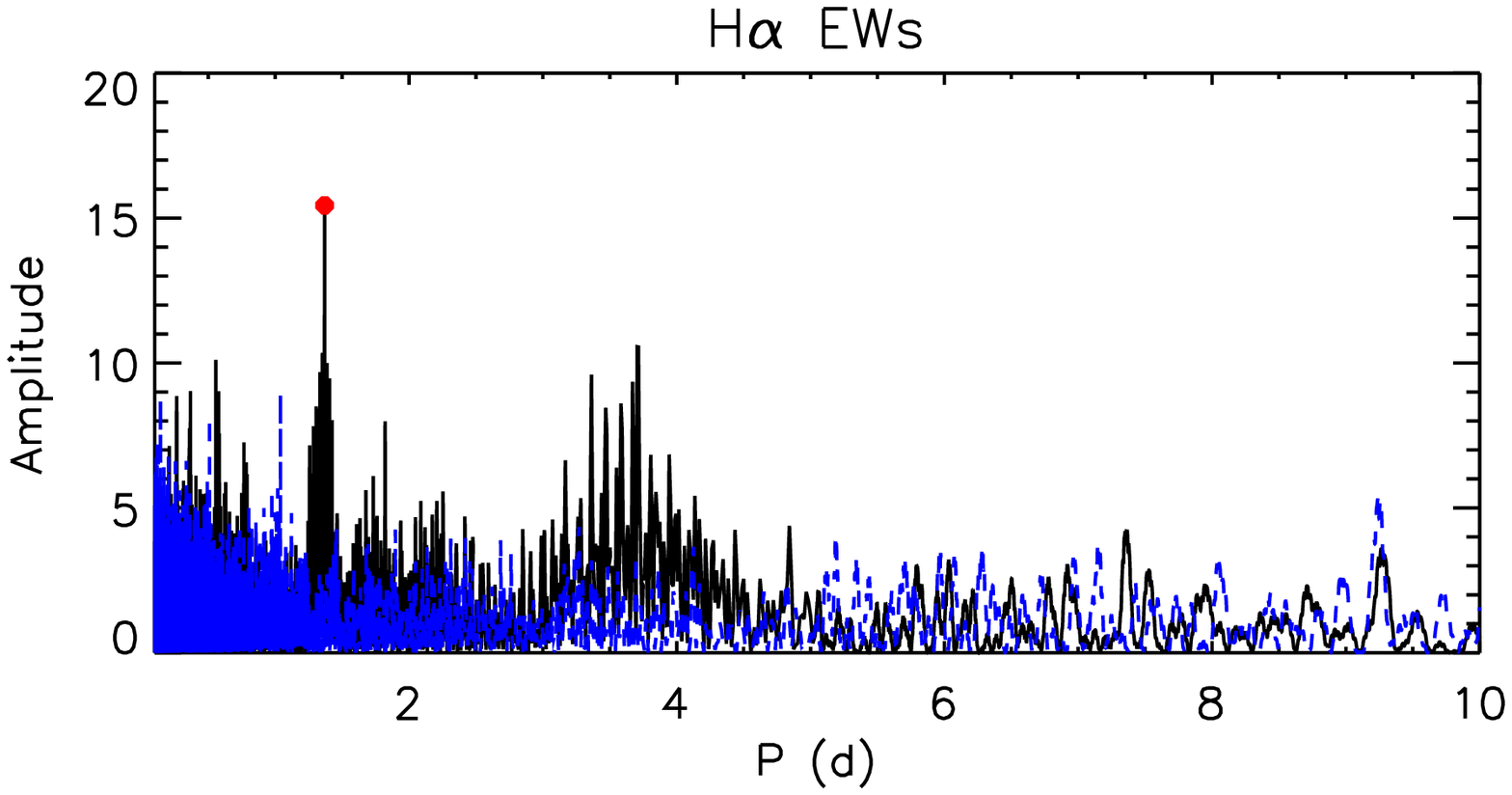}
\caption{Periodograms for (top panel): \bz~(solid black line) and \nz~(dashed blue), and (lower panel) H$\alpha$ EWs (where the dashed blue null periodogram was obtained from Gaussian noise with a standard deviation corresponding to the mean EW error bar). The maximum-amplitude period is indicated by a red circle.}
\label{bz_periods}
\end{figure}

\begin{figure}
\hspace{-0.25cm}\includegraphics[width=8.5cm]{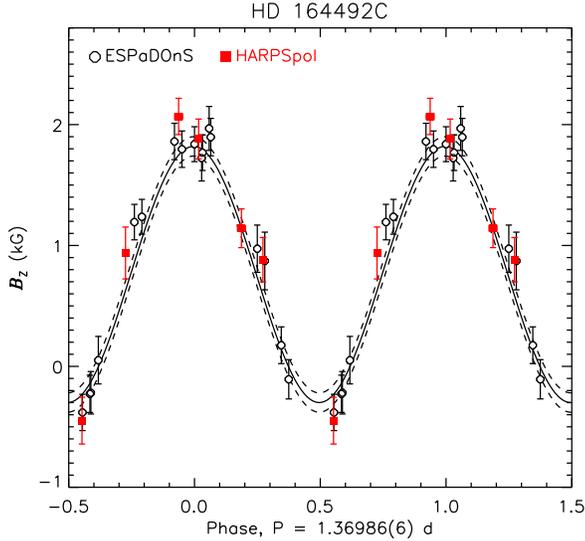}
\caption{\bz~(measured from LSD profiles extracted using He + metallic lines, disentangled as per Fig.\ \ref{lsd_stokesv}), phased according to Eq.~\ref{ephem}. Note the excellent agreement between the ESPaDOnS and HARPSpol measurements.}
\label{bz_curve}
\end{figure}

To determine the rotational period $P_{\rm rot}$, we performed Lomb-Scargle period analysis on the \bz~measurements. The periodogram is shown in the top panel of Fig.\ \ref{bz_periods}. There are two significant peaks in the periodogram at about 1.37~d and 3.64~d, with no corresponding peaks in the \nz~periodogram (indicated by dashed blue lines). As explored below, the star displays H$\alpha$ emission consistent with an origin in a circumstellar magnetosphere. The periodogram obtained from H$\alpha$ equivalent widths (EWs) is shown in the bottom panel of Fig.\ \ref{bz_periods}, with the strongest peak at 1.37~d. We reject the 3.64~d period based on two factors. First, the H$\alpha$ EWs phased with this period do not show a coherent variation (in agreement with the information provided by the periodogram) . Furthermore, if the \bz\ measurements are prewhitened with the 1.37~d period, the 3.64~d peak (as well as the most significant peaks below 1.0~d) are not longer present in the periodogram of those data. We therefore identify the shorter period as $P_{\rm rot}$. With the epoch JD0 defined at maximum \bz, we adopt the following ephemeris: 

\begin{equation}\label{ephem}
{\rm JD} = 2457197.93481 + (1.36986 \pm 0.00006)\cdot E,
\end{equation}

\noindent where the uncertainty in $P_{\rm rot}$ was determined using the formula from \cite{1976fats.book.....B}. The Stokes $V$ profiles in Fig.\ \ref{lsd_stokesv} are arranged in order of increasing rotational phase, and \bz~is shown phased according to Eq.(\ref{ephem}) in Fig.\ \ref{bz_curve}. We note that HARPSpol and ESPaDOnS measurements of the longitudinal field are in excellent mutual agreement.

\subsection{Physical parameters and surface magnetic field}\label{subsec_bzmodel}

The sinusoidal fit to the \bz~curve shown in Fig.\ \ref{bz_curve} has a reduced $\chi^2$ of 0.9, indicating that the variation of \bz~is well described by a sinusoid and, hence, that the surface magnetic field is dominated by the dipolar component, as is typical for magnetic, hot stars. The longitudinal magnetic field curve of a rotating dipole can be reproduced with the dipolar Oblique Rotator Model (ORM). This model is described by three parameters: the inclination $i$ of the stellar rotational axis from the line-of-sight, the obliquity angle $\beta$ of the magnetic axis from the rotational axis, and the surface magnetic field strength of the dipole at the magnetic pole $B_{\rm d}$. In this subsection we attempt to constrain these parameters.

Given $P_{\rm rot}$, the projected rotational velocity \vsini, and the stellar radius $R_*$, $i$ is given by 

\begin{equation}\label{inc}
\sin{i} = \frac{P_{\rm rot} v\sin{i}}{2\pi R_*}.
\end{equation}

The multiline fitting algorithm used to measure RVs included rotational and turbulent broadening, thus providing estimates of \vsini. From our earlier spectral modeling, we found that the broad-lined star C1 has \vsini~$=140\pm 10$ \kms, and the narrow-lined component C2 has \vsini~$=50\pm 5$~\kms. If we assume that $i=90^\circ$ and solve Eqn.\ \ref{inc} for $R_*$, the 1.37~d period {implies $R_1=3.8~R_\odot$}. For a \teff~of 26 kK, this implies a luminosity of $\log{L}=3.9$, and a log surface gravity of about 4.25\footnote{The discarded 3.64~d period would, on the other hand, require that $R_1=10.1~R_\odot$, a luminosity $\log L=4.7$, and $\log g=3.5$. This low gravity is not compatible with the modeling of the Balmer line profiles described earlier.}.

A more rigorous determination of the star's ORM parameters requires that we constrain $R_*$, which can be done if we know \teff~(determined in Sect.\ 3) and the luminosity $\log{L}$. In the following we attempt to determine the luminosities of the system's three components using constraints from distance, evolutionary models, and the presumed stellar ages. 

Although none of the HD 164492 members appear in the recent first Gaia data release (Gaia DR1), two other members of M20 (HD 164637 and HD 174492) are included. HD 164637 in particular is evaluated to have a membership probability of 99.6\% by \citet{2000A&AS..146..251B}. The Gaia parallaxes are respectively $0.782\pm 0.448$~mas ($1.28\pm 0.7$~kpc) and $0.836\pm 0.302$~mas ($1.20\pm 0.4$~kpc) \citep{2016arXiv160904303L}. Formally, both of these results are consistent with nearly the full range of published distances to the M20 cluster. To estimate the physical characteristics of the stars, we employ the weighted mean of the Gaia distances ($1.22\pm 0.35$~kpc), yielding a distance modulus ${\rm DM} = 10.4^{+0.6}_{-0.7}$~mag. The apparent visual magnitude of the system is $V=6.8$~mag.  Assuming no extinction $A_{\rm V}$ (given that the system is in a largely excavated part of the nebula), the absolute visual magnitude of the system is then $M_{\rm V} = V - A_{\rm V} - DM = -3.6 \pm 0.6$~mag. Using the bolometric correction $BC=-2.55\pm 0.21$ obtained via linear interpolation through the theoretical {\sc tlusty} BSTAR2006 grid \citep{Lanz2007} for \teff~$=26\pm2$~kK and $\log{g}=4.25\pm0.25$~(where $\log{g}<4$ is unlikely based on the spectral modelling), and assuming that $BC$ is about the same for the broad- and narrow-lined components, the system's absolute bolometric magnitude is $M_{\rm bol} = M_{\rm V} + BC = -6.2 \pm 0.8$~mag, yielding a total luminosity for the combined system of $\log{(L_{\rm sys}/L_\odot)} = (M_{\rm bol,\odot} - M_{\rm bol})/2.5 = 4.3 \pm 0.3$, where $M_{\rm bol,\odot}=4.74$~mag. Using the luminosity ratio $L_1/L_2 = 1.1\pm 0.1$ from Sect. 4, and assuming the contribution of C3 to the total luminosity is negligible, $\log{L_1/L_\odot} = 4.02 \pm 0.4$ and $\log{L_2/L_\odot} = 3.97 \pm 0.4$.  

\begin{figure}
\hspace{-0.4cm}\includegraphics[width=9cm]{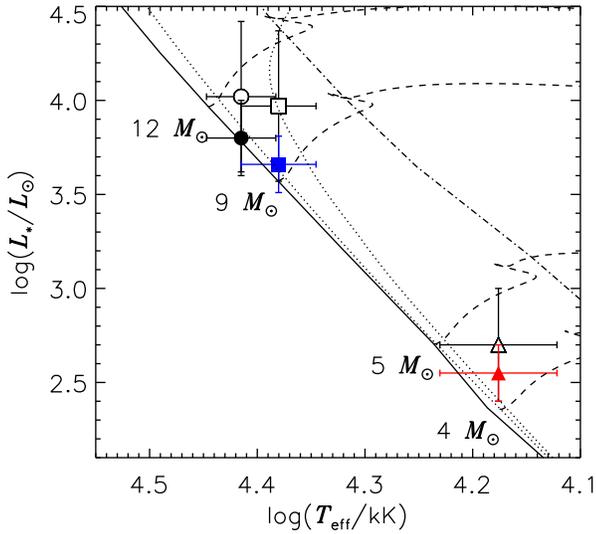}
\caption{The positions of the HD 164492C system's component stars on the \teff-$\log{L}$ diagram (Broad-lined C1 component, circle; narrow-lined C2 component, square; tertiary C3 component, triangle). Open symbols indicate photometrically determined values, filled symbols the final adopted values including constraints from orbital parameters and the age of the cluster. Evolutionary tracks from the \protect\cite{ekstrom2012} models are shown by dashed lines, while dotted lines indicate isochrones from the same models for $\log{t/{\rm yr}}=6.5$ and 7.2. The solid and dot-dashed lines indicate the ZAMS and TAMS, respectively. The luminosities are derived from photometry, however, given that the Trifid nebular is about 1 Myr old, all stars should be below the $\log{t}=6.5$ isochrone.}
\label{hrd_binary}
\end{figure}

The orbital properties of the close binary system offer an alternate means of constraining the luminosities of the component stars. The mass ratio of the C2 component and its companion C3 is $M_2/M_3 = K_3/K_2=2.08$. Assuming the stars are coeval, as should be the case for a close binary, they should be located on the same isochrone. We start from the photometrically determined luminosity of C2 of $\log{L_2/L_\odot} = 3.97 \pm 0.4$ and take the C2 component's mass to be $M_2 = 11 \pm 2~M_\odot$ from its position on the HRD. From the mass ratio this indicates that the C3 component must have a mass $M_3 = 5.3 \pm 1~M_\odot$. Tracing back along the $\log{t/{\rm yr}}=6.5$ {the youngest isochrone available in the grid, corresponding to a liberal upper limit on the age of the cluster} and $\log{t/{\rm yr}}=7.2$ isochrones (the maximum age for which both stars lie on the same isochrone) to obtain the luminosity of the C3 component yields $\log{(L_3/L_\odot)} = 2.57 \pm 0.24$. Requiring that the stars be coeval further restricts the C2 component's luminosity to $\log{(L_2/L_\odot)} = 3.8 \pm 0.3$, i.e. $\log{L_2} > 4.1$ is ruled out as the stars would no longer lie on the same isochrone. We therefore conclude, assuming that these stars are true, coeval members of the young cluster, that $L_2$ must be on the low end of the value inferred directly from the apparent magnitudes of the stars and the assumed distance (assuming that C2 and C3 lie on the same isochrone).

The open symbols in Fig.\ \ref{hrd_binary} show the positions of the three stars on the \teff-$\log{L}$ diagram, as inferred from spectral modelling, photometry, and assumed distance (for C1 and C2) and orbital properties (for C3). They are compared with the model evolutionary tracks of \citet{ekstrom2012}. Within the uncertainties in \teff~and $\log{L}$, the maximum age for which all three stars lie on the same isochrone is $\log{t/{\rm yr}}=7.2$, or about 16 Myr. However, this is much older than the reported age of the M20, which is estimated to be about $10^5$ yr \citep{1998Sci...282..462C}. If we restrict the luminosities of the stars such that they are located on the isochrone corresponding to this age, we obtain the positions corresponding to the filled symbols in Fig.\ \ref{hrd_binary}. If the age of the C1 component is restricted to be less than 1 Myr, then its luminosity must be $\log{(L_1/L_\odot)} = 3.8 \pm 0.15$. This would imply that the shorter estimates of the distance to the Trifid nebula (about 1 kpc) are likely to be more accurate, since they result in lower luminosities which locate all three stars on the correct isochrone. (This will be discussed further in Sect.~\ref{conclusion}.) This lower luminosity is adopted for the subsequent modeling of the star's magnetic field, yielding a stellar radius $R_1~=~\sqrt{(L_1/L_\odot)/(T_{\rm eff,1}/T_\odot)^4}~=~4.0 \pm 0.1 R_\odot$. The mass inferred from evolutionary models is $M_1 = 10.9 \pm 0.5 M_\odot$. As the spectroscopic modeling implies a luminosity ratio close to unity, this would then imply that the C2 component's luminosity must be similar ($\log{(L_2 / L_\odot)} = 3.65 \pm 0.15$), while the C3 component should be slightly lower than that inferred from purely photometric values ($\log{(L_3/L_\odot)}=2.56\pm~0.15$), yielding masses of $9.3\pm 0.4 M_\odot$ and $4.4 \pm 0.2 M_\odot$ for the C2 and C3 components, respectively. The corresponding luminosity ratios are $0.7 < L_1/L_2 < 2.8$ and $6 < L_2/L_3 < 25$, which are within the range of those determined via spectral modelling in Sect.\ 3. 

Using the radius $R_1$ determined from the C1 component's age-restricted luminosity, and Eq.\ \ref{inc}, $i = 63\pm 6\degr$. 

\cite{1967ApJ...150..547P} showed that $i$ and $\beta$ can be related by the parameter $r$, defined as

\begin{equation}\label{r}
r = \frac{|B_0| - B_1}{|B_0| + B_1} = \frac{\cos{(i + \beta)}}{\cos{(i - \beta)}},
\end{equation}

\noindent where $B_0$ and $B_1$ are obtained from the sinusoidal fit \bz$=B_0 + B_1\sin{2\pi\phi}$. The fit and its $1\sigma$ uncertainties are shown by the solid and dashed lines in Fig.\ \ref{bz_curve}, yielding $r=0.150\pm0.001$, thus $\beta =33 \pm 6$. The polar strength of the magnetic field is then \citep{1950MNRAS.110..395S}

\begin{equation}\label{bd_min}
B_{\rm d} = B_{z}^{\rm max}\left(\frac{15 + \mu}{20(3 - \mu)}(\cos{\beta}\cos{i} + \sin{\beta}\sin{i})\right)^{-1},
\end{equation}

\noindent where \bz$_{\rm max}=2.1\pm0.15$ kG is the maximum strength of the \bz~curve and $\mu$ is the linear limb-darkening coefficient, taken to be 0.37$\pm$0.02 via interpolation through the tables of \cite{diazcordoves1995} according to the star's \teff~and the surface gravity inferred from its position on the \teff-$\log{L}$ diagram. This yields $B_{\rm d} = 7.9^{+1.2}_{-1.0}$~kG

If the age restriction is not applied and the higher and more uncertain luminosity $\log{(L_1 / L_\odot)} = 4.02 \pm 0.4$ of the the C1 component is used instead, $R_1 = 5 \pm 1$ is also larger, yielding a smaller inclination~($i = 36^{+30\degr}_{-10}$), a larger obliquity~($\beta = 59^{+7\degr}_{-27}$), but essentially the same magnetic field strength ($B_{\rm d} = 7.1^{+0.7}_{-0.5}$~kG). Physical and magnetic parameters with and without age constraints are summarised in Table \ref{comp_params}.

\subsection{H$\alpha$ emission and variability}

 \begin{figure}
\hspace{-0.7cm} \includegraphics[width=9.4cm]{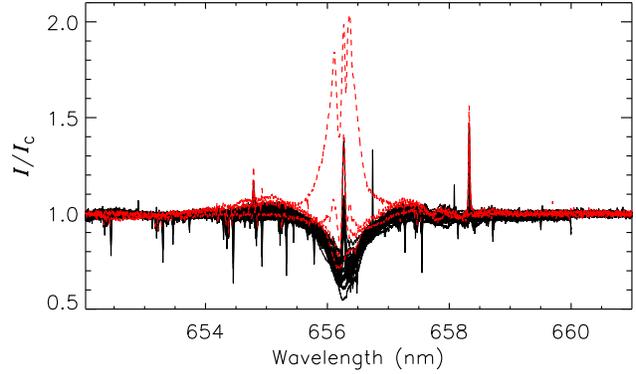}
 \caption{H$\alpha$ spectra from the ESPaDOnS, HARPSpol, FEROS, and UVES datasets. Observations rejected due to supposed contamination with light from the Herbig Be star HD164492D are indicated by dotted red lines; spectra included in the analysis are shown by solid black lines.}
 \label{halpha_oplot}
 \end{figure}

Fig.\ \ref{halpha_oplot} shows the H$\alpha$ line from all available spectra overplotted. The most apparent feature of these data are the sharp emission lines at 654.8 nm, 656.3 nm, and 658.5 nm: these are nebular in origin.

Five of the spectra in Fig.\ \ref{halpha_oplot} show emission that is qualitatively different from the others in strength and shape, with one of the FEROS spectra in particular showing particularly strong emission. These spectra are highlighted with red dotted lines. These data do not phase coherently with $P_{\rm rot}$, and all but one of these spectra show metallic and He lines that differ from the much larger collection of remaining spectra. In two cases, one FEROS and one UVES spectrum, normalization appears to have failed, leading to excess emission in the wings. In the remaining cases the emission pattern is very different from that exhibited by other magnetic B stars, with very strong emission in the inner part of the line. The two ESPaDOnS spectra showing this pattern both have the highest seeing in the ESPaDOnS dataset: 1.68", as compared to a mean of 1.2"$\pm$0.1. Given the crowded field in the vicinity of this star, it is likely that this anomalous emission is due to contamination from the nearby Herbig Be star HD 164492D (separated from HD 164492C by 2"). These data were consequently ignored in the remainder of the H$\alpha$ analysis, but they will be discussed again at the end of the paper. 

\begin{figure}
\includegraphics[width=8cm]{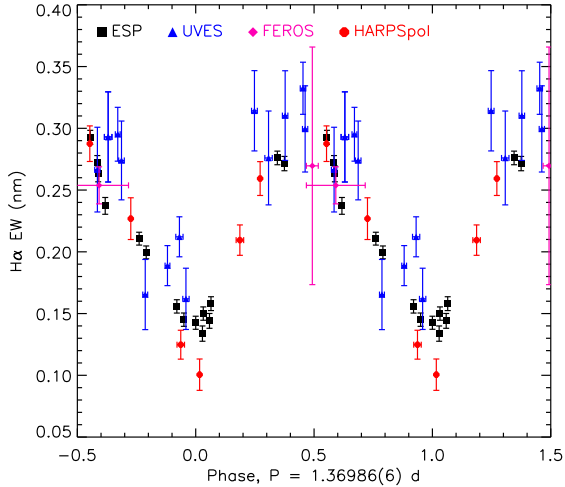}
\caption{H$\alpha$ EW measurements phased according to Eq.~\ref{ephem}.}
\label{halpha_ew}
\end{figure}

The remaining spectra all show enhanced emission in the line wings, similar to that seen in other rapidly rotating, strongly-magnetized early B-type stars. If this emission is associated with a magnetosphere, it would be expected to phase coherently with $P_{\rm rot}$, with maximum emission corresponding to an extremum of \bz. As discussed in Sect.~\ref{LSD}, the Lomb-Scargle periodogram of these data  shows a single strong peak, consistent with the period $P_{\rm rot}$ derived from the magnetic measurements. Fig.\ \ref{halpha_ew} shows the H$\alpha$ EWs phased with Eq.~\ref{ephem}, demonstrating that there is a coherent variation. Maximum emission occurs at phase 0 (maximum \bz), and minimum emission occurs at phase 0.5 (when \bz~is closest to 0 G). The single-wave variation of the H$\alpha$ EW is furthermore expected given that, according to the observed longitudinal field variation, the only one magnetic pole of the star is visible during a rotational cycle.

\begin{figure}
\includegraphics[width=9cm]{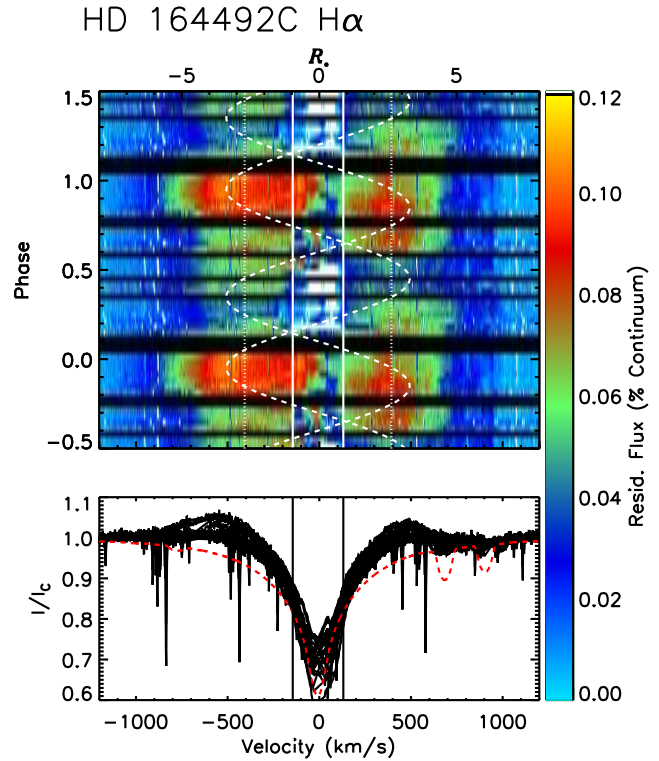}
\caption{H$\alpha$ dynamic spectrum, phased according to Eq. 2. Residual flux is obtained by subtracting synthetic spectra from the observed spectra, where the synthetic spectra use the physical parameters determined in Section 6, adjusted to the RVs measured in Section 5. Nebular emission was removed from observed spectra by linear interpolation between pixels adjacent to the emission lines; telluric lines were removed with a low-pass filter. The bottom panel shows the individual line profiles, with a representative synthetic spectrum, created with RV~$=0$~\kms~for both components, shown by the dashed red line. In the top panel, residual flux is mapped to the colour-bar (right) and plotted as a function of rotational phase. Solid vertical lines indicate $\pm$\vsini, and dotted vertical lines indicate $\pm$\rk. Dashed lines trace out a two-cloud model of the magnetosphere.}
\label{halpha_dyn}
\end{figure}

The periodic variability of the H$\alpha$ emission, locked in phase with the longitudinal magnetic field variation, provides strong evidence for the presence of hot plasma held in corotation with the star by its magnetic field.

A dynamic spectrum of the H$\alpha$ line is shown in Fig.\ \ref{halpha_dyn}. Residual flux was obtained by subtracting synthetic from observed spectra, where the model spectra were created using the physical parameters for the broad-lined and narrow-lined components from Sect.~3, shifted to the appropriate RVs from Table~\ref{RV_table}. For display purposes, nebular emission was removed by linear interpolation between pixels adjacent to each nebular emission line. Telluric absorption lines were removed with a low-pass filter. The emission behaves as expected for a centrifugal magnetosphere (CM; \citealt{petit2013}), peaking at high ($\sim 4v\sin i$) velocities, and increasing and decreasing roughly symmetrically in the blue and red line wings. 

The emission appears to extend across the full line profile. This is as expected if $\beta$ is small, in which case the Rigidly Rotating Magnetosphere model (RRM; \citealt{2005MNRAS.357..251T}) predicts a slightly warped disk with only mild density enhancements at the intersections of the rotational and magnetic equators. This behaviour is consistent with the geometry derived using the age-constrained physical parameters. Interpreting variability in the rotationally broadened core (inside the vertical solid lines) is complicated by the nebular emission and residuals from the binary motion, thus whether or not there are periodic eclipses cannot be evaluated. 

The blue-shifted emission is somewhat stronger (and possible more extended) than the red-shifted emission. This could indicate an asymmetry in the magnetosphere, possibly arising from a surface magnetic field topology somewhat more complex than a dipole \citep[e.g.][]{2015MNRAS.451.2015O}. The curved dashed lines in Fig.\ \ref{halpha_dyn} trace out a simple two-cloud model, with velocity maxima aligned with the blue- and red-shifted emission peaks. These appear to be offset by about 0.1 in rotational phase. A dipolar RRM model predicts blue- and red-shifted emission maxima occurring at the same phase, thus this could, in addition to the apparent asymmetry in emission strength, be another indication of contributions to the surface magnetic field from higher-order multipoles. 

\subsection{Magnetospheric parameters}

The radial extent of magnetic confinement of the wind is given by the Alfv\'en radius. This is defined as the point at which the Alfv\'enic Mach number (the ratio of the wind velocity to the Alfv\'en velocity) is equal to unity. This in turn scales with the magnetic wind confinement parameter $\eta_*$ defined by \cite{2002ApJ...576..413U} as the balance between the energy density of the magnetic field and the kinetic energy density of the stellar wind. Using the \cite{2001A&A...369..574V} mass-loss prescription and the age-constrained stellar parameters, the mass-loss rate is $\log{\dot{M}/(M_\odot/{\rm yr})}=-9.3\pm0.3$, and the wind terminal velocity \vinf$=2600\pm30$ \kms. Solving equation 7 from \cite{2002ApJ...576..413U} using these wind parameters and $B_{\rm d}=7.9^{+1.2}_{-1.0}$~kG and $R_*=4.0\pm0.1 R_\odot$ as in Sect.\ \ref{subsec_bzmodel} yields $\eta_*=4^{+8}_{-3}\times 10^4$ at the magnetic equator, where the resistance of the magnetic field to the stellar wind is at a maximum. Using equation 8 from \cite{2008MNRAS.385...97U} the equatorial Alfv\'en radius is then \ra$=17\pm3 R_*$.

In rapidly rotating stars, rotational support of the circumstellar plasma plays a key role in shaping the magnetosphere, both by enabling accumulation via prevention of gravitational infall, and by compressing the plasma into a geometrically thin, warped disk following the effective potential minima \citep{2005MNRAS.357..251T}. The corotation or Kepler radius \rk, at which the gravitational and centrifugal forces balance, is found to be $R_{\rm K}=2.8\pm0.1 R_*$ from the stellar mass and rotational frequency using equation 12 from \cite{2005MNRAS.357..251T}. In Fig.\ \ref{halpha_dyn}, $\pm$\rk~is indicated by vertical dotted lines, and coincides closely with the emission maxima. This is consistent with the magnetically confined plasma being located above \rk. 

A dimensional parameterization of the magnetospheric volume is given by $\log{(R_{\rm A}/R_{\rm K})} = 0.80\pm0.08$. This is on the high end for magnetic B-type stars, and is consistent with the ratio observed for the subset of stars with H$\alpha$-bright CMs \citep{petit2013}. 

In Section \ref{subsec_bzmodel} the luminosity was constrained by the presumed maximum age of the system. If this assumption is relaxed, the results of the current section are largely unchanged: $\log{\dot{M}} = -9.0 \pm 0.3$, $v_\infty = 2200 \pm 200$~\kms, $\eta_*=(5\pm 3)\times 10^4$, \ra$=15.9\pm1.8~R_*$, \rk$=1.9 \pm 0.4~R_*$, and $\log{(R_{\rm A}/R_{\rm K})} = 0.9\pm0.1$. The fundamental conclusion is that \ra$\gg$\rk. The largest difference is in \rk, which is smaller due to the larger radius and, consequently, the higher surface equatorial rotational velocity $v_{\rm eq} = 190 \pm 30$~\kms, as compared to $v_{\rm eq}=154\pm7$ \kms~with the age-restricted parameters. Comparing to Fig.\ \ref{halpha_dyn}, this would place \rk~at a much lower radius than the emission peaks. 

HD 164492C has similar magnetospheric and rotational parameters to other magnetic B-type emission-line stars, so it is not unreasonable to interpret its emission in the context of the RRM. RRM makes the assumption that $\eta_*\gg 1$ and, hence, the wind plasma will be entirely locked to the magnetic field in the vicinity of the emission line-forming region. The three-dimensional distribution of the plasma then follows an accumulation surface defined by the minima of the effective potential, assuming hydrostatic equilibrium along each field line. For an aligned rotator ($\beta=0$) this yields a disk in the magnetic and equatorial planes. Larger angles of $\beta$ introduce a warp to the disk, with the two densest regions at the intersections of the magnetic and rotational equators. For large $\beta$ this results in the formation of two distinct clouds. 

Examining Fig.\ \ref{halpha_dyn}, we note that there is emission present at all phases. The strongest emission occurs at phase 0, coincident with the maximum of the magnetic field: at this phase, the magnetic pole is closest to the centre of the projected stellar disk, thus the magnetosphere is at quadrature and has its greatest projected area. At phase 0.5, \bz$\sim0$ G, indicating that the magnetic equator is bisecting the projected stellar disk. The projected area of the magnetosphere is now at its minimum, as is the H$\alpha$ emission strength. This modulation in emission strength only makes sense if the plasma is fairly dense, such that it is optically thick at some or all rotational phases. This is a general characteristic of stars with emission-line CMs. It is instructive to compare HD 164492C to similar stars with well-defined magnetic and rotational parameters. Broadly speaking, the CM emission is either confined within two distinct red and blue clouds and present at only some rotational phases (e.g. $\sigma$ Ori E, HR 7355, HD 176582, \citealt{2015MNRAS.451.2015O,2013MNRAS.429..177R,2011AJ....141..169B}) or apparent across the line profile and present at all rotational phases, indicative of plasma distributed more evenly at all azimuths (e.g. $\delta$ Ori C, HR 5907, HD 23478,  \citealt{2010MNRAS.401.2739L,2012MNRAS.419.1610G,2015MNRAS.451.1928S}). In the former case $\beta$ is generally large (above 45$^\circ$), while amongst the latter stars $\beta$ is generally small (below about 15$^\circ$). The presence of emission at all phases and velocities is more consistent with the relatively small $\beta=33\pm6^\circ$ determined using age-restricted parameters than it is with the larger $\beta=59^{+7}_{-27}\degr$ inferred from the unrestricted parameters. 

\subsection{Rotational evolution}

The extended moment arm of the rigidly rotating magnetosphere serves as an efficient braking mechanism \citep{1967ApJ...148..217W,2009MNRAS.392.1022U}. The spin-down timescale $\tau_{\rm J}$ for HD 164492C is, using equation 8 from \cite{2009MNRAS.392.1022U}, about $12\pm3$~Myr with age-restricted parameters, or $3.7\pm$2 Myr using unrestricted parameters, where the longer spindown timescale is due to the lower \mdot~inferred for a younger, less luminous star.

In the age-restricted case $\tau_{\rm J}$ is an order of magnitude longer than the cluster age of about 1 Myr; without age restrictions $\tau_{\rm J}$ is only about twice the cluster age, but is much less than the age of the star that would be inferred from evolutionary models, 12.1$\pm$3.5 Myr. 

This could imply that the star did not start on the ZAMS at critical rotation. The rotation parameter $W\equiv v_{\rm eq}/v_{\rm orb}$ \citep{2008MNRAS.385...97U}, where $v_{\rm orb}$ is the velocity necessary to maintain a Keplerian orbit at the equatorial surface. When $W=1$ the star is rotating at its critical velocity. Using equation 11 from \cite{2008MNRAS.385...97U}, $W=0.32\pm0.09$ with unrestricted parameters, or $0.20\pm0.03$ with age-restriction. Taking $W_0 = We^{t/\tau_{\rm J}}$ and using the 1 Myr cluster age yields $W_0=0.22$ in the age-restricted case and $W_0=0.48$ in the unrestricted case. It is unknown if the star experienced any braking on the pre-main sequence, however, this result would suggest that $W_0$ cannot have been much larger than its current value. This conclusion depends on the mass-loss rate: if \mdot$\sim$1.5 dex higher than determined using Vink mass-loss, $\tau_{\rm J}$ would be of a similar magnitude to $t$ and hence compatible with $W_0=1$. 

\cite{petit2013} calculated maximum spindown ages $t_{\rm S,max}$ under the assumption that $W_0=1$. Assuming a gyration radius $r_{\rm gyr}= 0.3 R_*$ (as used by \citealt{petit2013}), and using age-restricted parameters, we obtain $t_{\rm S,max} = 21 \pm 7$~Myr. If instead we use $r_{\rm gyr} = 0.25 R_*$ (which is closer to the value predicted by the stellar structure models of \citealt{claret2004}), we obtain $t_{\rm S,max}=12\pm$5~Myr.  The spindown age can be brought into agreement with the cluster age of $\sim$1 Myr if $r_{\rm gyr} \sim 0.1 R_*$. If instead the unrestricted parameters are used, $t_{\rm S,max}=7\pm3$~Myr with $r_{\rm gyr} = 0.3~R_*$, and $4\pm2$~Myr with $r_{\rm gyr} = 0.25~R_*$.

Thus the cluster and gyrochronological ages can be reconciled by: 1) using unrestricted parameters (but this assumes stellar parameters that are themselves inconsistent with the stars' effective temperatures and ages); 2) making $r_{\rm gyr}$ much smaller than predicted by models of unmagnetized stars; or 3) by assuming that the star experienced significant braking on the pre-main sequence, such that $W_0 < 1$.

We note that if we instead adopt the theoretical wind model of \citet{2014A&A...564A..70K}, differences in our conclusions are relatively small. This is because the effective temperature of the magnetic star is fairly high and in this regime the Krticka mass-loss rates are not terribly different from those of Vink. The value of $\dot M$ is about the same, but that of $v_\infty$ is about 1000 km/s higher, resulting in a smaller \ra\ (about $10~R_*$) but about the same spindown timescale. Moreover, there is no difference between the age restricted/unrestricted values, since the Krticka $\dot M$ values don't vary as significantly  as the star evolves (the increase due to higher luminosity is compensated by the decrease due to the lower $T_{\rm eff}$).†

\section{Discussion and conclusion}\label{conclusion}

\begin{table}
\begin{centering}\label{comp_params}
\begin{tabular}{lrr}
\hline
Quantity & value ($t < 1$ Myr) & value (unrestricted age) \\
\hline
{\bf HD 164492C1}\\
\hline
$v\sin i$ (km/s) & \multicolumn{2}{c}{$140\pm 10$} \\
$P_{\rm rot}$ (d) & \multicolumn{2}{c}{1.36986$\pm$0.00007} \\
JD$0$ & \multicolumn{2}{c}{2457197.93481} \\
$T_{\rm eff}$ (kK) & \multicolumn{2}{c}{26$\pm$2} \\
$\log{L}$ ($L_\odot$) & 3.8$\pm$0.15 & 4.02$\pm$0.4 \\
$M$ ($M_\odot$) & 10.9$\pm$0.5 & 11$\pm$2 \\
$R$ ($R_\odot$) & 4.0$\pm$0.1 & 5$\pm$1 \\
$B_{\rm d}$ (kG) & 7.9$^{+1.2}_{-1.0}$ & 7.1$^{+0.7}_{-0.5}$ \\
$i$ ($\degr$) & 63$\pm$6 & $36^{+30}_{-10}$ \\
$\beta$ ($\degr$) & 33$\pm$6  & $59^{+7}_{-27}$ \\
\hline
{\bf HD 164492C2}\\
\hline
$v\sin i$ (km/s) & \multicolumn{2}{c}{$50\pm 5$} \\
$T_{\rm eff}$ (kK) & \multicolumn{2}{c}{24$\pm$2} \\
$\log{L}$ ($L_\odot$) & 3.65$\pm$0.15 & 3.97$\pm$0.4 \\
$M$ ($M_\odot$) & 9.3$\pm$0.4 & 10$\pm$1 \\
$R$ ($R_\odot$) & 3.6$\pm$0.1 & 6$\pm$1.5 \\
\hline
{\bf HD 164492C3}\\
\hline
$v\sin i$ (km/s) & \multicolumn{2}{c}{30} \\
$T_{\rm eff}$ (kK) & \multicolumn{2}{c}{15$\pm$2} \\
$\log{L}$ ($L_\odot$) & 2.56$\pm$0.15 & 2.7$\pm$0.24 \\
$M$ ($M_\odot$) & 4.4$\pm$0.2 & 4.5$\pm$0.3 \\
$R$ ($R_\odot$) & 2.4$\pm$0.1 & 3.3$\pm$0.6 \\
\hline\hline
\end{tabular}
\caption{Physical, rotational and magnetic properties of the components used and derived in this study. Parameters with two columns correspond to those derived with the age restricted to less than 1 Myr (left) and those derived with no restriction on the age (right).}
\end{centering}
\end{table}

In this paper we used high resolution spectroscopy and spectropolarimetry to derive the physical properties and magnetic characteristics of the multiple system HD\,164492C. In addition to the previously-known broad-lined and narrow-lined early B stars (components HD\,164492C1 and C2), we also identified a late B-type companion (HD\,164492C3). The physical properties assumed and derived for these three stars are summarised in Table~\ref{comp_params}.

Components C2 and C3 exhibit significant radial velocity (RV) variations. We modeled these variations as orbital motion about a common centre-of-mass, with a period of $P_{\rm orb}=12.5351(7)$~d and eccentricity $e=0.515(6)$. The magnetic C1 component exhibits no discernible radial velocity variations with this period, or any other period shorter than the span of our data.

We derived the physical properties of the 3 components. When the stars are constrained to lie on isochrones consistent with the reported age of M20, we find masses of $10.9\pm 0.5~M_\odot$ (C1), $9.3\pm 0.4~M_\odot$ (C2) and $4.4\pm 0.2~M_\odot$ (C2). All three stars are inferred to be main sequence objects. If we use the distance to M20 implied by the Gaia parallax measurements, we obtained somewhat larger masses for all 3 stars. This results in older implied ages of the 3 stars: the error boxes are consistent with isochrones as old as $\log t=7.2$, or about 16~Myr.   

From both measurements of the variable longitudinal magnetic field and the variation of the equivalent width of strong, high velocity emission in the H$\alpha$ line, we infer the rotation period of the magnetic star $P_{\rm rot}=1.36986(6)$~d. We derive the star's magnetic geometry, finding $i=63\pm 6\degr$, $\beta=33\pm 6\degr$ and a dipole polar strength $B_{\rm d}=7.9^{+1.2}_{-1.0}$~kG from the age-restricted parameters. From the unrestricted parameters, we obtained a different geometry ($i=36^{+30}_{-10}\degr$ and $\beta=59^{+7}_{-27}\degr$), but a similar field strength $B_{\rm d}=7.1^{+0.7}_{-0.5}$~kG). (Although the inferred masses and luminosities derived using the two approaches described above agree within their uncertainties, they ultimately result in radii of the C1 star, and hence magnetic geometries, that are very different from one another.) We interpret the variable H$\alpha$ emission as originating in a centrifugal magnetosphere surrounding the C1 component. Based on the behaviour of the H$\alpha$ emission, we are led to prefer the model with smaller derived $\beta$ (i.e. that implied by the age-restricted parameters) as most consistent with our observations of HD\,164492C.

Having derived the evolutionary masses of the components, we can compare with the projected dynamical masses of the C2 and C3 components to infer the inclination of the binary orbit. For both stars, the projected dynamical masses in combination with the age-constrained evolutionary masses imply $i=54\degr$. This allows us to deproject the semimajor axis, yielding $a=0.25$~AU. At periastron, the two components are separated by $a(1-e)\simeq 0.12$~AU, or about 7 times the radius of component C2. 

The total luminosity of the system implied by the individual luminosities of the 3 stars under the constraint $t < 1$~Myr is $\log{(L/L_\odot)} =  4.04 \pm 0.15$, and assuming that C1 and C2 have the same bolometric correction, then the apparent $V$ magnitude of the system would require a distance $d = 0.9 \pm 0.2$~kpc. This calculation assumes that the extinction $A_V$ is negligible. If $A_V$ is nonzero, then the distance must be even smaller, e.g. $A_V = 0.25$ would lead to the stars being about 100 pc closer. This provides new evidence that the distance to the Trifid nebula corresponds to the smaller values reported in the literature. This distance determination, which combines constraints from the C2/C3 system's orbital properties and the reported age of the star cluster, is smaller than (but still formally consistent with) the Gaia parallax distance.

Upon examining the complete collection of ESPaDOnS, UVES, FEROS and HARPSpol spectral available for this system, we found that a small number of spectra exhibited H$\alpha$ emission profiles that were qualitatively different from the remainder. We concluded that these spectra were contaminated by light from nearby stars, in particular HD\, 164492D. It is instructive to note that we also attempted to model the spectral energy distribution of the system using broadband photometry in the optical and near-infrared, as well as IUE spectroscopy. We found large inconsistencies between the various epoch of IUE data, and between the IUE data and the optical photometry. Ultimately, we were unable to usefully interpret those data, and we view their behaviour as further evidence for contamination of some spectra and spectrophotometric data by the light of nearby stars.

During our modeling of the combined spectrum of the system, we observed that many of the He~{\sc i} lines (e.g. He~{\sc i} $\lambda 4144$, He~{\sc i} $\lambda 4388$, and He~{\sc i} $\lambda 4471$) showed broad wings that were not reproduced by the model. We suspect that this is evidence for the presence of strong horizontal or vertical He abundance non-uniformities in the atmosphere of the magnetic C1 star. However, in our analysis we found no clear evidence for strong variability of the EWs of He lines. A more sophisticated analysis of the spectra - likely including a complete disentangling of significant spectra regions - is required to better understand the origin of this phenomenon.

The relationship of the magnetic C1 star to the binary system remains an outstanding question. That a self-consistent set of physical parameters for the three stars can be obtained assuming a common distance and age suggests that they are not very far separated in distance or age.  Given the high stellar density in the central region of M20, the association of the binary with C1 might be a line-of-sight conjunction of relatively close by physically unassociated stars. On the other hand, the magnetic star could be physically associated with the C2+C3 binary in a wider, longer period orbit. An examination of the RVs of C1 over the 1.9 years that our observations span shows no significant, systematic change in the seasonal mean velocities. As a consequence, the existing data do not allow us to choose between these two hypotheses. Multi-epoch high resolution imaging or long-term spectroscopic monitoring would be needed to say more.

 \section*{Acknowledgments}
We acknowledge the contribution of A. Tkachenko. GAW acknowledges Discovery Grant support from the Natural Sciences and Engineering Research Council of Canada. We acknowledge financial support from "le Programme National de Physique Stellaire" (PNPS) of CNRS/INSU, France. EA acknowledges financial support from "le Laboratoire d'Excellence OSUG@2020 (ANR10 LABX56)" funded by "le programme d'Investissements d'Avenir". This research has made use of the SIMBAD database, of the VizieR catalogue access tool, and of "Aladin sky atlas", all operated at CDS, Strasbourg, France.

\bibliography{article-arxiv}

\end{document}